\documentclass[preprint,
amsmath,amssymb,aps,nofootinbib, Letter]{revtex4-1}
\usepackage{graphicx}
\usepackage{dcolumn}
\usepackage{bm}
\usepackage{color}

\usepackage{geometry} 
\usepackage{fancyhdr} 
\usepackage{moresize} 

\usepackage{multirow}
\usepackage{longtable}
\usepackage{xcolor}
\usepackage{colortbl,booktabs}

\newcommand{\la}{\langle}
\newcommand{\ra}{\rangle}
\newcommand{\qbar}{\bar{q}}
\newcommand{\ubar}{\bar{u}}

\newcommand{\sbar}{\overline{s}}
\newcommand{\cbar}{\overline{c}}

\def\B{{\cal B}}

\def\up{\uparrow}
\def\dw{\downarrow}
\def\be{\begin{eqnarray}}
\def\en{\end{eqnarray}}

\begin{document}


\title{Two-body weak decays of doubly charmed baryons }

\author{Hai-Yang Cheng}
\affiliation{%
 Institute of Physics, Academia Sinica, Taipei, Taiwan 115, Republic of China }
 %

\author{Guanbao Meng, Fanrong Xu\footnote{fanrongxu@jnu.edu.cn}, Jinqi Zou}
\affiliation{
 Department of Physics, Jinan University,
 Guangzhou 510632, People's Republic of China
}%


\bigskip
\begin{abstract}
\bigskip
The hadronic two-body weak decays of the doubly charmed baryons $\Xi_{cc}^{++}, \Xi_{cc}^+$
and $\Omega_{cc}^+$ are studied in this work. To estimate the nonfactorizable contributions, we work in the pole model for the $P$-wave amplitudes and current algebra for $S$-wave ones.
For the $\Xi_{cc}^{++}\to \Xi_c^+\pi^+$ mode, we find a large destructive interference between factorizable and nonfactorizable contributions for both $S$- and $P$-wave amplitudes. Our prediction of $\sim 0.70\%$ for its branching fraction is smaller than the earlier estimates in which nonfactorizable effects were not considered, but agrees nicely with the result based on an entirely different approach, namely, the covariant confined quark model. On the contrary, a large constructive interference was found in the $P$-wave amplitude by Dhir and Sharma, leading to a branching fraction of order $(7-16)\%$.
Using the current results for the absolute branching fractions of $(\Lambda_c^+,\Xi_c^+)\to p K^-\pi^+$ and the LHCb measurement of $\Xi_{cc}^{++}\to\Xi_c^+\pi^+$ relative to $\Xi_{cc}^{++}\to\Lambda_c^+ K^- \pi^+\pi^+$,
we obtain $\B(\Xi_{cc}^{++}\to\Xi_c^+\pi^+)_{\rm expt}\approx (1.83\pm1.01)\%$ after employing the latest prediction of $\B(\Xi_{cc}^{++}\to\Sigma_c^{++}\overline{K}^{*0})$.
Our prediction of $\mathcal{B}(\Xi_{cc}^{++}\to\Xi_c^+\pi^+)\approx 0.7\%$ is thus consistent with the experimental value but in the lower end. It is important to pin down the branching fraction of this mode in future study.
Factorizable and nonfactorizable $S$-wave amplitudes interfere constructively in $\Xi_{cc}^+\to\Xi_c^0\pi^+$. Its large branching fraction of order 4\% may enable experimentalists to search for the $\Xi_{cc}^+$ through this mode. That is, the $\Xi_{cc}^+$ is reconstructed through the $\Xi_{cc}^+\to\Xi_c^0\pi^+$ followed by the decay chain $\Xi_c^0\to \Xi^-\pi^+\to p\pi^-\pi^-\pi^+$. Besides $\Xi_{cc}^+\to\Xi_c^0\pi^+$, the $\Xi_{cc}^+\to\Xi_c^+ (\pi^0,\eta)$ modes also receive large nonfactorizable contributions to their $S$-wave amplitudes. Hence, they have large branching fractions among $\Xi_{cc}^+\to \B_c+P$ channels.
Nonfactorizable amplitudes in $\Xi_{cc}^{++}\to \Xi_c^{'+}\pi^+$  and $\Omega_{cc}^+\to \Xi_c^{'+}\overline{K}^0$ are very small compared to the factorizable ones owing to the Pati-Woo theorem for the inner $W$-emission amplitude. Likewise, nonfactorizable $S$-wave amplitudes in $\Xi_{cc}^+\to\Xi_c^{'+}(\pi^0,\eta)$ decays are also suppressed by the same mechanism.

\end{abstract}

\maketitle


\small
\section{Introduction}\label{sec:Intro}

The doubly charmed baryon state $\Xi_{cc}^{++}$ was first discovered by the LHCb in the weak decay mode $\Lambda_c^+ K^-\pi^+\pi^+$  \cite{Aaij:2017ueg} and subsequently confirmed in another mode $\Xi_c^+\pi^+$ \cite{Aaij:2018gfl}. Its lifetime  was also measured by the LHCb to be \cite{Aaij:2018wzf}
\begin{equation} \label{eq:lifetime}
\tau_{\Xi_{cc}^{++}}=0.256^{+0.024}_{-0.022}({\rm{stat.}})\pm 0.014({\rm{syst.}}) \,{\rm{ps}}.
\end{equation}
The updated mass is given by \cite{Aaij:2019uaz}
\begin{equation} \label{eq:Xiccmass}
m_{\Xi_{cc}^{++}}=3621.55\pm0.23\pm0.30~{\rm MeV}.
\end{equation}

As the first two-body weak decay $\Xi_c^+ \pi^+$ of the doubly charmed baryon $\Xi_{cc}^{++}$ was reported by the LHCb with the result \cite{Aaij:2018gfl}
\begin{equation} \label{eq:Xicpi}
\frac{\mathcal{B}(\Xi_{cc}^{++}\to\Xi_c^+\pi^+)\times \mathcal{B}(\Xi_c^+\to p K^-\pi^+)}
{\mathcal{B}(\Xi_{cc}^{++}\to\Lambda_c^+ K^- \pi^+\pi^+)\times
\mathcal{B}(\Lambda_c^+\to p K^- \pi^+)}=0.035\pm 0.009 ({\rm{stat.}})
\pm 0.003({\rm{syst.}}),
\end{equation}
we would like to investigate in this work the nonleptonic two-body decays of doubly charmed baryons $\Xi_{cc}^{++}$, $\Xi_{cc}^{+}$ and $\Omega_{cc}^+$. This has been studied intensively in the literature \cite{Li:2017ndo,Yu:2017zst,Wang:2017mqp,Wang:2017azm,Sharma:2017txj,Gutsche:2017hux,Dhir:2018twm,Jiang:2018oak,Gutsche:2018msz,Shi:2019hbf,Gerasimov:2019jwp,Zhao:2018mrg,Gutsche:2019iac,Ke:2019lcf}. Many authors \cite{Wang:2017mqp,Shi:2019hbf,Gerasimov:2019jwp,Ke:2019lcf} considered only
the factorizable contributions from the external $W$-emission governed by the Wilson coefficient $a_1$. It is well known that in charmed baryon decays, nonfactorizable contributions from $W$-exchange or inner $W$-emission diagrams play an essential role and they cannot be neglected, in contrast with the negligible effects in heavy meson decays. Unlike the meson case, $W$-exchange is no longer subject to helicity and color suppression. The
experimental measurements of the decays $\Lambda_c^+\to\Sigma^0\pi^+,~ \Sigma^+\pi^0$ and
$\Xi^0K^+$, which do not receive any factorizable
contributions, indicate that $W$-exchange and inner $W$-emission indeed are important in charmed baryon decays. By the same token, it is expected that nonfactorizable contributions are also important in doubly charmed baryon decays.

In the 1990s various approaches were developed to describe the nonfactorizable effects in hadronic decays of singly charmed baryons $\Lambda_c^+$, $\Xi_c^{+,0}$ and $\Omega_c^0$. These include the covariant confined quark model \cite{Korner:1992wi,Ivanov:1997ra}, the pole model \cite{Xu:1992vc,Cheng:1991sn,Cheng:1993gf,Zenczykowski:1993jm} and
current algebra \cite{Cheng:1993gf,Sharma:1998rd}. In the same vein, some of these techniques have been applied to the study of $W$-exchange in doubly charmed baryon decays. For example,
$W$-exchange  contributions to the $P$-wave amplitude were estimated by Dhir and Sharma \cite{Sharma:2017txj,Dhir:2018twm} using the pole model. However, nonfactorizable corrections to the $S$-wave amplitudes were not addressed by them. Likewise, Long-distance effects due to $W$-exchange have been estimated in \cite{Gutsche:2018msz,Gutsche:2017hux,Gutsche:2019iac} within the framework of the covariant confined quark model.
Long-distance contributions due to $W$-exchange or inner $W$-emission were modeled as final-state rescattering effects in \cite{Yu:2017zst,Jiang:2018oak}. This approach has been applied to $\B_{cc}\to\B_c V$ ($V$: vector meson) \cite{Jiang:2018oak}.

In the pole model, nonfactorizable $S$- and $P$-wave amplitudes for $1/2^+\to 1/2^++0^-$ decays are dominated by $1/2^-$ low-lying baryon resonances and $1/2^+$ ground-state baryon poles, respectively. However,
the estimation of pole amplitudes is a difficult and nontrivial
task since it involves weak baryon matrix elements and strong
coupling constants of ${1\over 2}^+$ and ${1\over 2}^-$ baryon
states.  As a consequence, the evaluation of pole diagrams is far more uncertain
than the factorizable terms. This is the case in particular for $S$-wave terms as
they require the information of the troublesome negative-parity baryon resonances which are not well understood in the quark model. This is the main reason why the nonfactorizable $S$-wave amplitudes of doubly charmed baryon decays were not considered in
\cite{Sharma:2017txj,Dhir:2018twm} within the pole model.

It is well known that the pole model  is reduced to current algebra for $S$-wave amplitudes in the soft pseudoscalar-meson limit. In the soft-meson limit, the intermediate excited $1/2^-$ states in the $S$-wave amplitude can be summed up and reduced to a
commutator term. Using the relation $[Q_5^a, H_{\rm eff}^{\rm PV}]=-[Q^a, H_{\rm eff}^{\rm PC}]$, the parity-violating (PV) amplitude is simplified to a simple commutator term expressed in terms of parity-conserving (PC) matrix elements. Therefore, the great advantage of current algebra is that the evaluation of the parity-violating $S$-wave amplitude does not require the information of the negative-parity $1/2^-$ poles.  Although the pseudoscalar meson produced in $\B_c\to \B+P$ decays is in general not truly soft, current algebra seems to work empirically well for $\Lambda_c^+\to \B+P$ decays \cite{Cheng:2018hwl,Zou:2019kzq}. Moreover, the predicted negative decay asymmetries by current algebra for both $\Lambda_c^+\to \Sigma^+\pi^0$ and $\Sigma^0\pi^+$ agree in sign with the recent BESIII measurements \cite{Ablikim:2019zwe} (see \cite{Cheng:2018hwl,Zou:2019kzq} for details). In contrast, the pole model or the covariant quark model and its variant always leads to a positive decay asymmetry for aforementioned two modes. Therefore,
in this work we shall follow \cite{Cheng:2018hwl,Zou:2019kzq} to work out
the nonfactorizable $S$-wave amplitudes in doubly charmed baryon decays using current algebra and the $W$-exchange contributions to $P$-wave ones using the pole model.

In short, there exist three entirely distinct approaches for tackling the nonfactorizable contributions in doubly charmed baryon decays: the covariant confined quark model (CCQM) , final-state rescattering and the pole model in conjunction with current algebra. As stressed in \cite{Gutsche:2018msz,Gutsche:2017hux,Gutsche:2019iac}, the evaluation of the $W$-exchange diagrams in CCQM is technically quite demanding since it involves a three-loop calculation. The calculation of triangle diagrams for final-state rescattering is also rather tedious.
Among these different analyses, current algebra plus the pole model turns out to be the simplest one.

Since the decay rates and decay asymmetries are sensitive to the relative sign between factorizable and non-factorizable amplitudes, it is important to evaluate all the unknown parameters in the model in a globally consistent convention to ensure the correctness of their relative signs once the wave function convention is fixed. In our framework, there are three important quantities: form factors, baryonic matrix elements and axial-vector form factors.
All of them will be evaluated in the MIT bag model. We shall see later that the branching fractions of $\Xi_{cc}^{++}\to\Xi_c^+\pi^+$ and $\Xi_{cc}^+\to\Xi_c^0\pi^+$ modes are  quite sensitive to their interference patterns.

This paper is organized as follows.  In Sec. II we set up the framework for the analysis of
hadronic weak decays of doubly charmed baryons, including the topological diagrams and
the formalism for describing factorizable and nonfactorizable terms. We present the explicit expressions of nonfactorizable amplitudes for both $S$- and $P$-waves.  Baryon matrix elements and axial-vector form factors calculated in the MIT bag model are also summarized.
Numerical results and discussions are presented in Sec. III.
A conclusion will be given in Sec. \ref{sec:con}.
In the Appendix, we write down the doubly charmed baryon wave functions to fix our convention.

\section{Theoretical  framework}

\begin{figure}[t]
\begin{center}
\includegraphics[width=0.80\textwidth]{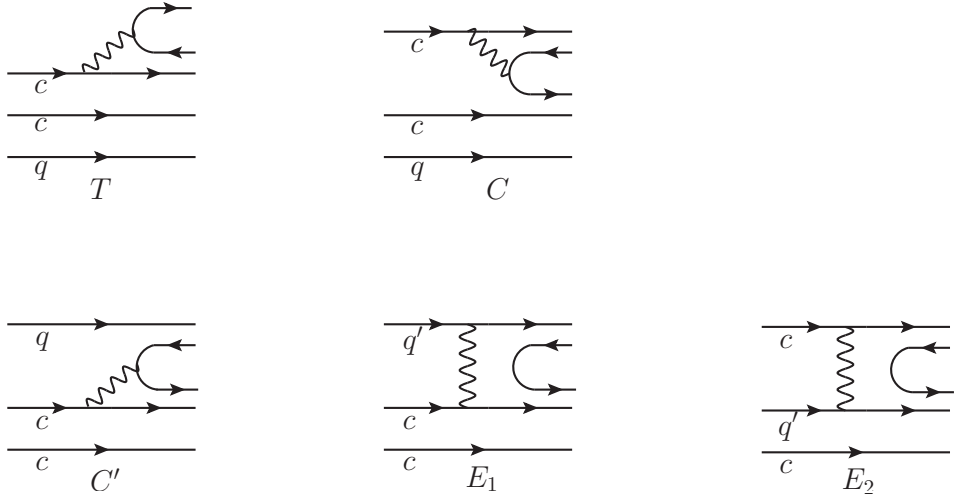}
\vspace{0.1cm}
\caption{Topological diagrams contributing to $\B_{cc}\to \B_c+P$ decays: external $W$-emission $T$, internal $W$-emission $C$, inner $W$-emission $C'$,  $W$-exchange diagrams $E_1$ and $E_2$, where $q=u,d,s$ and $q'=d,s$.} \label{fig:Bicc}
\end{center}
\end{figure}

In this work we shall follow \cite{Cheng:1991sn,Cheng:1993gf} closely with many quantities and operators well defined in these references. 

\subsection{Topological diagrams}
More than two decades ago,  Chau, Tseng and one of us (HYC) have presented
a general formulation of the topological-diagram scheme for the nonleptonic weak decays of baryons \cite{Chau:1995gk}, which was then applied to all the decays of the antitriplet and sextet charmed baryons. For the weak decays $\B_{cc}\to B_c+P$ of interest in this work, the relevant topological diagrams are
the external $W$-emission $T$, the internal $W$-emission $C$, the inner $W$-emission $C'$,  and the $W$-exchange diagrams $E_1$ as well as $E_2$ as depicted in Fig. \ref{fig:Bicc}. Among them, $T$ and $C$ are factorizable, while $C'$ and $W$-exchange give nonfactorizable contributions.  The relevant topological diagrams  for all Cabibbo-favored  decay modes of doubly charmed baryons  are shown  in Table \ref{tab:modes}.

We notice from Table \ref{tab:modes} that (i) there are two purely factorizable modes: $\Xi_{cc}^{++}\to \Sigma_c^{++}\overline{K}^0$ and $\Omega_{cc}^+\to\Omega_c^0\pi^+$, (ii) the $W$-exchange contribution manifests only in $\Xi_{cc}^+$ decays, and (iii)
the topological amplitude $C'$ in $\Xi_{cc}^{++}\to \Xi_c^{'+}\pi^+$, $\Xi_{cc}^+\to \Xi_c^{'+}(\pi^0,\eta)$ and $\Omega_{cc}^+\to\Xi_c^{'+}\overline{K}^0$ should vanish
because of the Pati-Woo theorem \cite{Pati:1970fg} which results from the facts that the $(V-A)\times(V-A)$ structure of weak interactions is invariant under the Fierz transformation and that the baryon wave function is color antisymmetric. This theorem requires that the quark
pair in a baryon produced by weak interactions be antisymmetric in flavor. Since the sextet
$\Xi'_c$ is symmetric in light quark flavor, it cannot contribute to $C'$. We shall see below that this feature is indeed confirmed in realistic calculations.

\begin{table}[t]
\caption{Topological diagrams contributing to two-body Cabibbo-favored decays of the doubly charmed baryons $\Xi_{cc}^{++}, \Xi_{cc}^+$
and $\Omega_{cc}^+$.
\label{tab:modes}}
\begin{ruledtabular}
\begin{tabular}{l l l l l l}
$\Xi_{cc}^{++}$ &  Contributions  & $\Xi_{cc}^+$ & Contributions & $\Omega_{cc}^+$ & Contributions \\
\colrule
$\Xi_{cc}^{++} \to \Sigma_c^{++} \overline{K}^0$ & $C$ &  $ \Xi_{cc}^+\to\Xi_c^0 \pi^+$  & $T,E_1$  & $ \Omega_{cc}^+ \to \Omega_c^0 \pi^+ $ & $T$ \\
$\Xi_{cc}^{++} \to \Xi_{c}^+ \pi^+$ & $T,C'$   &  $ \Xi_{cc}^+\to\Xi_c^{'0} \pi^+$ & $T,E_1$ &   $ \Omega_{cc}^+ \to \Xi_c^+ \overline{K}^0 $ & $C,C'$ \\
$\Xi_{cc}^{++} \to \Xi_{c}^{'+} \pi^+$    & $T,C'$ & $ \Xi_{cc}^+\to\Lambda_c^{+} \overline{K}^0$ &  $C,E_2$ &   $ \Omega_{cc}^+ \to \Xi_c^{'+} \overline{K}^0 $ & $C,C' $\\
& $$ & $ \Xi_{cc}^+\to\Sigma_c^{+} \overline{K}^0$  & $C,E_2$ \\
& $$ & $ \Xi_{cc}^+\to\Xi_c^{+} \pi^0$  & $C',E_1$  \\
& $$ & $ \Xi_{cc}^+\to\Xi_c^{'+} \pi^0$  & $C',E_1$ \\
& $$ & $ \Xi_{cc}^+\to\Xi_c^{+} \eta$  & $C',E_1,E_2$  \\
& $$ & $ \Xi_{cc}^+\to\Xi_c^{'+} \eta$  & $C',E_1,E_2$  \\
& $$ & $ \Xi_{cc}^+\to\Sigma_c^{++} K^-$  & $E_2$  \\
& $$ & $ \Xi_{cc}^+\to\Omega_c^{0} K^+$  & $E_1$  \\
\end{tabular}
\end{ruledtabular}
\end{table}

\subsection{Kinematics}\label{sec:Kin}
The amplitude for two-body weak decay $\mathcal{B}_i\to\mathcal{B}_f P$ is given as
\begin{equation}
M(\mathcal{B}_i \to \mathcal{B}_f P)= i\ubar_f (A-B\gamma_5) u_i,
\end{equation}
where $B_i (B_f)$ is the initial (final) baryon and $P$ is a pseudoscalar meson.
The decay width and up-down decay asymmetry are given by
\begin{align}
& \Gamma=\frac{p_c}{8\pi}
\left[ \frac{(m_i+m_f)^2-m_P^2}{m_i^2}|A|^2
+\frac{(m_i-m_f)^2-m_P^2}{m_i^2}|B|^2\right], \nonumber \\
& \alpha=\frac{2\kappa {\rm{Re}} (A^* B)}{|A|^2+\kappa^2|B|^2},
\end{align}
where $p_c$ is the three-momentum
in the rest frame of the mother particle and $\kappa=p_c/(E_f+m_f)=\sqrt{(E_f-m_f)/(E_f+m_f)}$. The $S$- and $P$- wave amplitudes of the two-body decay generally receive both factorizable and non-factorizable contributions
\begin{align}
& A=A^{\rm{fac}}+A^{\rm{nf}},\quad
B=B^{\rm{fac}}+B^{\rm{nf}}.
\label{eq:amplitude}
\end{align}

\subsection{Factorizable amplitudes}\label{sec:fac}

The description of the factorizable contributions of the doubly charmed baryon decay $\mathcal{B}_{cc}\to\mathcal{B}_c P$
is based on the effective Hamiltonian approach. In the following we will
give explicitly the factorizable contribution of $S$- and $P$-wave amplitudes.

The effective Hamiltonian for the Cabibbo-favored process reads
\begin{align} \label{eq:effH}
&\mathcal{H}_{\rm{eff}}=\frac{G_F}{\sqrt{2}}V_{cs}V_{ud}^*(c_1O_1+c_2O_2)+h.c., \\
&O_1=(\sbar c)(\ubar d),\quad
O_2=(\ubar c)(\sbar d),\qquad (\qbar_1 q_2)\equiv \qbar_1\gamma_\mu(1-\gamma_5) q_2, \nonumber
\end{align}
where $c_1$ and $c_2$ are Wilson coefficients.
Under the factorization hypothesis 
the amplitude can be written as
\begin{equation}
M=\la P\mathcal{B}_c|\mathcal{H}_{\rm{eff}}|\mathcal{B}_{cc}\ra
=\left\{\begin{array}{ll}
\frac{G_F}{\sqrt{2}}V_{cs}V_{ud}^* a_{1} \la P|(\ubar d)|0\ra \la \mathcal{B}_c|(\sbar c)|\mathcal{B}_{cc}\ra , &P =\pi^+, \\ \\
\frac{G_F}{\sqrt{2}}V_{cs}V_{ud}^* a_{2} \la P|(\sbar d)|0\ra \la \mathcal{B}_c|(\ubar c)|\mathcal{B}_{cc}\ra, &P=\overline{K}^0,
\end{array}
\label{eq:CF}
\right.
\end{equation}
where $a_1=c_1+\frac{c_2}{N_c}, a_2=c_2+\frac{c_1}{N_c}$.
One-body and two-body matrix elements of the current are parameterized in terms of decay constants and form factors, respectively,
\begin{equation}
\la K (q)|\sbar\gamma_\mu(1-\gamma_5) d|0\ra = if_K q_\mu,\quad
\la \pi (q)|\ubar\gamma_\mu(1-\gamma_5) d|0\ra = if_\pi q_\mu,
\label{eq:KFF}
\end{equation}
with $f_\pi=132$ MeV, $f_K=160$ MeV and
\begin{eqnarray}  \label{eq:FF}
\la \mathcal{B}_c(p_2)|\cbar\gamma_\mu(1-\gamma_5) u|\mathcal{B}_{cc}(p_1)\ra
&=&\ubar_2 \left[ f_1(q^2) \gamma_\mu -f_2(q^2)i\sigma_{\mu\nu}\frac{q^\nu}{M}+f_3(q^2)\frac{q_\mu}{M}\right. \nonumber \\
&&\hspace{0.5cm} -\left.\left(g_1(q^2)\gamma_\mu-g_2 (q^2)i\sigma_{\mu\nu}\frac{q^\nu}{M}+g_3(q^2)
\frac{q_\mu}{M}
\right)\gamma_5
\right]u_1,
\end{eqnarray}
with the  initial particle mass  $M$ and the momentum transfer  $q=p_1-p_2$.
Then the factorizable amplitude has the expression
\begin{equation}
M(\mathcal{B}_{cc}\to \mathcal{B}_c P)
=i\frac{G_F}{\sqrt{2}}a_{1,2} V_{ud}^*V_{cs} f_P \ubar_2(p_2)\left[(m_1-m_2)f_1(q^2)
+(m_1+m_2)g_1(q^2) \gamma_5\right]u_1(p_1),
\end{equation}
where we have neglected the contributions from the form factors $f_{3}$ and $g_{3}$.\footnote{To see the possible corrections from the form factors $f_3$ and $g_3$ for kaon or $\eta$ production in the final state, we notice that $m_P^2/m_{\Lambda_c}^2=0.047$ for the kaon and 0.057 for the $\eta$. Since the form factor $f_3$ is much smaller than $f_1$ (see e.g. Table IV of \cite{Zhao:2018zcb}), while $g_3$ is of the same order as $g_1$, it follows that the form factor $f_3$ can be safely neglected in the factorizable amplitude, while $g_3$ could make $\sim5\%$ corrections for kaon or $\eta$ production. For simplicity,  we will drop all the contributions from $f_3$ and $g_3$.
} 

Hence,
\begin{align}
&A^{\rm{fac}}=\frac{G_F}{\sqrt{2}}a_{1,2} V_{ud}^*V_{cs} f_P(m_{\mathcal{B}_{cc}}-m_{\mathcal{B}_c}) f_1(q^2),
\nonumber\\
&B^{\rm{fac}}= -\frac{G_F}{\sqrt{2}}a_{1,2} V_{ud}^*V_{cs} f_P(m_{\mathcal{B}_{cc}}+m_{\mathcal{B}_c})
g_1(q^2).
\end{align}

There are two different non-perturbative parameters in the factorizable amplitudes: the decay constant and the form factor. Unlike the decay constant, which can be measured directly by experiment, the form factor is less known experimentally.
Form factors defined in Eq. (\ref{eq:FF}) have been evaluated in various models: the MIT bag model \cite{PerezMarcial:1989yh}, the non-relativistic quark model \cite{PerezMarcial:1989yh}, heavy quark effective theory \cite{Cheng:1995fe},  the light-front quark model \cite{Wang:2017mqp,Ke:2019lcf} and light-cone sum rules \cite{Shi:2019hbf}.

\begin{table}[t]
\linespread{1.5}
\footnotesize{
 \caption{The calculated form factors with $c\to s$ transition in the MIT bag model at maximum four-momentum
 transfer squared $q^2=q^2_{\rm{max}}=(m_i-m_f)^2$ and at $q^2=m_P^2$. 
 } \label{tab:FF1}
 \vspace{0.3cm}
\begin{center}
\begin{tabular}
{ cccr|ccr}
 \hline\hline
 Modes &  $f_1(q_{\rm{max}}^2)$ &$ f_1(m_P^2)/f_1(q_{\rm{max}}^2)$ & $f_1(m_P^2)$ & $g_1(q_{\rm{max}}^2)$ &
 $g_1(m_P^2)/g_1(q_{\rm{max}}^2)$  & $g_1(m_P^2)$ \\
   \midrule\hline
$\Xi_{cc}^{++}\to\Sigma_c^{++} \overline{K}^0$   & $Y_1$  &  $0.540$ &  $0.476$~ & $\frac53 Y_2$  & $0.673$ & $0.862$    \\
$\Xi_{cc}^{++}\to\Xi_c^{+} \pi^+$  & $\frac{\sqrt{6}}{2}Y_1^s$  &  $0.496$  &$0.577$~    & $\frac{\sqrt{6}}{6}Y_2^s$  &      $0.634$  &  $0.222$  \\
$\Xi_{cc}^{++}\to\Xi_c^{'+} \pi^+$   & $\frac{\sqrt{2}}{2}Y_1^s$  &$0.575$    &$0.386$~    & $\frac{5\sqrt{2}}{6}Y_2^s$  &      $0.695$&   $0.703$   \\
\midrule\hline
$\Xi_{cc}^{+}\to\Lambda_c^{+} \overline{K}^0$ & $\frac{\sqrt{6}}{2}Y_1$  & $0.487$  &    $0.526$~      & $\frac{\sqrt{6}}{6}Y_2$  &  $0.632$  &  $0.198$       \\
$\Xi_{cc}^{+}\to\Sigma_c^{+} \overline{K}^0$   & $\frac{\sqrt{2}}{2}Y_1$  &  $0.622$  &      $0.388$~                &      $\frac{5\sqrt{2}}{6}Y_2$  &  $0.734$  &  $0.665$    \\
$\Xi_{cc}^{+}\to\Xi_c^{0} \pi^+$  & $\frac{\sqrt{6}}{2}Y_1^s$  &$0.572$    &     $0.666$~         &
$\frac{\sqrt{6}}{6}Y_2^s$  &
$0.693$      &   $0.243$    \\
$\Xi_{cc}^{+}\to\Xi_c^{'0} \pi^+$  & $\frac{\sqrt{2}}{2}Y_1^s$  & $0.648$   &       $0.435$~             &
$\frac{5\sqrt{2}}{6}Y_2^s$  &      $0.749$  &  $0.758$       \\
\midrule\hline
$\Omega_{cc}^{+}\to\Omega_c^{0} \pi^+$   & $Y_1^s$  &  $0.532$  &   $0.505$~          &
$\frac{5}{3}Y_2^s$  & $0.661$ &   $0.947$           \\
$\Omega_{cc}^{+}\to\Xi_c^{+} \overline{K}^0$   & $-\frac{\sqrt{6}}{2}Y_1$  & $0.406$   &   $-0.438$~       & $-\frac{\sqrt{6}}{6}Y_2$  & $0.568$  &    $-0.178$       \\
$\Omega_{cc}^{+}\to\Xi_c^{'+} \overline{K}^0$   & $\frac{\sqrt{2}}{2}Y_1$  & $0.495$   &   $0.309$~       &
$\frac{5\sqrt{2}}{6}Y_2$  &  $0.638$ &  $0.578$   \\
\hline\hline
\end{tabular}
\end{center}
 }
\end{table}

In this work we shall follow the assumption of nearest pole dominance \cite{Fakirov:1977ta} to write down the $q^2$ dependence of form factors as
\begin{equation}
f_i(q^2)=\frac{f_i(0)}{(1-q^2/m_V^2)^2},\qquad
g_i(q^2)=\frac{g_i(0)}{(1-q^2/m_A^2)^2},
\label{eq:FFq2}
\end{equation}
where $m_V=2.01\,{\rm GeV}$, $m_A=2.42\,{\rm GeV}$ for the $(c\bar{d})$ quark content, and
$m_V=2.11\,{\rm GeV}$, $m_A=2.54\,{\rm GeV}$ for the $(c\bar{s})$ quark content.
In the zero recoil limit where $q^2_{\rm{max}}=(m_i-m_f)^2$, the form factors are expressed in the MIT bag model to be \cite{Cheng:1993gf}
\begin{align}
& f_1^{\mathcal{B}_f \mathcal{B}_i}(q^2_{\rm{max}})=\la \mathcal{B}_f\uparrow| b_{q_1}^\dagger
b_{q_2}| \mathcal{B}_i\uparrow\ra \int d^3 \bm{r}(u_{q_1}(r)u_{q_2}(r)+v_{q_1}(r)v_{q_2}(r)),
\nonumber \\
& g_1^{\mathcal{B}_f \mathcal{B}_i}(q^2_{\rm{max}})=\la \mathcal{B}_f\uparrow| b_{q_1}^\dagger b_{q_2} \sigma_z| \mathcal{B}_i\uparrow\ra \int d^3 \bm{r}
(u_{q_1}(r)u_{q_2}(r)-\frac13v_{q_1}(r)v_{q_2}(r)),
\end{align}
where $u(r)$ and $v(r)$ are the large and small components, respectively, of the quark wave function in the bag model. Form factors  at different $q^2$ are related by
\begin{equation}
f_i(q_2^2)=\frac{(1-q_1^2/m_V^2)^2}{(1-q_2^2/m_V^2)^2}f_i(q_1^2),\qquad
g_i(q_2^2)=\frac{(1-q_1^2/m_A^2)^2}{(1-q_2^2/m_A^2)^2}g_i(q_1^2).
\label{eq:FF2}
\end{equation}
Numerical results of the form factors at $q^2=m_\pi^2$ for various $\B_{cc}\to \B_c$ transitions are shown in Table \ref{tab:FF1}. In the calculation we have defined the bag integrals
\begin{align}
&Y_1=4\pi \int r^2 dr (u_u u_c+v_u v_c)=0.8825,\qquad Y_1^s=4\pi \int  r^2 dr (u_s u_c+v_s v_c)=0.9500, \nonumber \\
&Y_2=4\pi \int r^2 dr (u_u u_c-\frac13 v_u v_c)=0.7686,\quad~ Y_2^s=4\pi \int  r^2  dr (u_s u_c-\frac13 v_s v_c)=0.8588.
\end{align}

In Table III we compare the form factors evaluated in the MIT bag model with the recent calculations based on the light-front quark model (LFQM) \cite{Wang:2017mqp,Ke:2019lcf} and light-cone sum rules (QSR) \cite{Shi:2019hbf}. There are two different
LFQM calculations denoted by LFQM(I) \cite{Wang:2017mqp} and LFQM(II) \cite{Ke:2019lcf}, respectively. They differ in the inner structure of $\B_{cc}\to \B_c$ transition: a quark-diquark picture of charmed baryons in the former and a three-quark picture in the latter. We see from Table \ref{tab:FFcomp} that form factors are in general largest in LFQM(I) and smallest in QSR.

\begin{table}[t]
\footnotesize{
\begin{center}
\caption{Form factors $f_1(q^2)$ and $g_1(q^2)$ at $q^2=m_\pi^2$ for various $\B_{cc}\to \B_c$ transitions evaluated in the MIT bag model,  the light-front quark models, LFQM(I) \cite{Wang:2017mqp}  and LFQM(II)  \cite{Ke:2019lcf}, and QCD sum rules (QSR) \cite{Shi:2019hbf}.
} \label{tab:FFcomp}
 \vspace{0.3cm}
\begin{tabular}{l *4{c}| *4{c}}
\toprule\hline
\multirow{2}{*}{${\cal B}_{cc}\to {\cal B}_c$}    &
\multicolumn{4}{c}{$f_1(m_\pi^2)$ } &
\multicolumn{4}{c}{$g_1(m_\pi^2)$ } \\
\cline{2-5} \cline{6-9}
  & ~~MIT~~ & ~~LFQM(I)~~ & ~~LFQM(II)~~ &~~QSR~~   & ~~MIT~~ &~~LFQM(I)~~ & ~~LFQM(II)~~ & ~~QSR~~   \\
\midrule\hline
  $\Xi_{cc}^{++}\to \Xi_c^{+}$     &  0.577 & 0.920 & 0.734 & 0.664
 & 0.222 & 0.259 & 0.172 & 0.095   \\
 $\Xi_{cc}^{++}\to \Xi_c^{'+}$    &  0.386  & 0.541 & 0.407 & 0.360 & 0.703 & 0.731 & 0.496 & 0.208
 \\
  $\Xi_{cc}^{+}\to \Xi_c^{0}$     &  0.606 & 0.920 & 0.734 & 0.664
 & 0.243 & 0.259 & 0.172 & 0.095   \\
  $\Xi_{cc}^{+}\to \Xi_c^{'0}$    &  0.435  & 0.541 & 0.407 & 0.360 & 0.758 & 0.731 & 0.496 & 0.208
 \\
$\Omega_{cc}^{+}\to \Omega_c^{0}$     &  0.505 & 0.758 & & 0.420
 & 0.947 & 1.025 & & 0.150   \\
\bottomrule\hline
\end{tabular}
\end{center}}
\end{table}

\subsection{Nonfactorizable amplitudes}\label{sec:nf}
We shall adopt the pole model to describe the nonfactorizable contributions.
The general formulas for $A$ ($S$-wave) and $B$ ($P$-wave) terms in the pole model are given by
\begin{align}
& A^{\rm{pole}}=-\sum\limits_{B_n^*(1/2^-)}\left[\frac{g_{_{B_f B_n^* P}}\,b_{n^* i}}{m_i-m_{n^*}} +
\frac{b_{fn^*}\,g_{_{B_{n}^*B_i P}}}{m_f-m_{n^*}}\right],   \nonumber\\
& B^{\rm{pole}}=\sum\limits_{B_n}\left[ \frac{g_{_{B_f B_n P}}\,a_{ni}}{m_i-m_n}
+\frac{a_{fn}\,g_{_{B_n B_i P}}}{m_f-m_n}\right],
\label{eq:pole}
\end{align}
with the baryonic matrix elements
\begin{equation} \label{eq:m.e.}
\la \mathcal{B}_n|H|\mathcal{B}_i\ra = \ubar_n (a_{ni} + b_{n i}\gamma_5) u_i,\qquad
\la \mathcal{B}_i^*(1/2^-) | H|\mathcal{B}_j\ra =\ubar_{i*} b_{i^* j} u_j.
\end{equation}
It is known that the estimate of the $S$-wave amplitudes in the pole model is a difficult and nontrivial task as it involves the matrix elements and strong coupling constants of $1/2^-$ baryon resonances which we know very little \cite{Cheng:1991sn}. \footnote{Attempts of explicit calculations of intermediate $1/2^-$ pole contributions to the $S$-wave amplitudes had been made before in \cite{Cheng:1991sn,Cheng:1993gf}.
}
Nevertheless, if the emitted pseudoscalar meson is soft, then the intermediate excited baryons can be summed up, leading to a commutator term
\be
A^{\rm{com}} &=& -\frac{\sqrt{2}}{f_{P^a}}\la \mathcal{B}_f|[Q_5^a, H_{\rm{eff}}^{\rm PV}]|\mathcal{B}_i\ra
=\frac{\sqrt{2}}{f_{P^a}}\la \mathcal{B}_f|[Q^a, H_{\rm{eff}}^{\rm PC}]|\mathcal{B}_i\ra, \label{eq:Apole}
\en
with
\begin{equation}
Q^a=\int d^3x \qbar\gamma^0\frac{\lambda^a}{2}q,\qquad
Q^a_5=\int d^3x \qbar\gamma^0\gamma_5\frac{\lambda^a}{2}q.
\end{equation}
Likewise, the $P$-wave amplitude is reduced in the soft-meson limit to
\be
B^{\rm{ca}} &=& \frac{\sqrt{2}}{f_{P^a}}\sum_{\mathcal{B}_n}\left[ g^A_{\mathcal{B}_f \mathcal{B}_n}\frac{m_f+m_n}{m_i-m_n}a_{ni}
+a_{fn}\frac{m_i + m_n}{m_f-m_n} g_{\mathcal{B}_n \mathcal{B}_i}^A\right],
\label{eq:Bpole}
\en
where the superscript ``ca" stands for current algebra and 
we have applied the generalized Goldberger-Treiman relation
\begin{equation} \label{eq:GT}
g_{_{\mathcal{B'B}P^a}}=\frac{\sqrt{2}}{f_{P^a}}(m_{\mathcal{B}}+m_{\mathcal{B'}})g^A_{\mathcal{B'B}}.
\end{equation}
In Eq. (\ref{eq:Bpole}) $a_{ij}$ is the parity-conserving matrix element defined in Eq. (\ref{eq:m.e.}) and $g_{ij}^A$ is the axial-vector form factor defined in Eq. (\ref{eq:GT}).
Eqs. (\ref{eq:Apole}) and (\ref{eq:Bpole}) are the master equations for nonfactorizable amplitudes in the pole model under the soft meson approximation.

\subsubsection{$S$-wave amplitudes}
\label{subsec:Swave}

As shown in Eq. (\ref{eq:Apole}), the nonfactorizable $S$-wave amplitude is determined by the commutator terms of conserving charge $Q^a$ and the parity-conserving part of the effective Hamiltonian $H_{\rm{eff}}^{\rm PC}$.
Below we list the $A^{\rm com}$ terms for various meson production:
\begin{align} \label{eq:commu}
&A^{\rm{com}}(\mathcal{B}_i\to \mathcal{B}_f \pi^{\pm})=\frac{1}{f_\pi}\la \mathcal{B}_f|[I_{\mp}, H_{\rm{eff}}^{\rm PC}]|\mathcal{B}_i\ra,  \nonumber\\
&A^{\rm{com}}(\mathcal{B}_i\to \mathcal{B}_f \pi^{0})=\frac{\sqrt{2}}{f_\pi}\la \mathcal{B}_f|[I_3, H_{\rm{eff}}^{\rm PC}]|\mathcal{B}_i\ra,  \nonumber\\
&A^{\rm{com}}(\mathcal{B}_i\to \mathcal{B}_f \eta_8)=\sqrt{\frac32}\frac{1}{f_{\eta_8}}\la \mathcal{B}_f|[Y, H_{\rm{eff}}^{\rm PC}]|\mathcal{B}_i\ra,   \\
&A^{\rm{com}}(\mathcal{B}_i\to \mathcal{B}_f K^{\pm})=\frac{1}{f_K}\la \mathcal{B}_f|[V_{\mp}, H_{\rm{eff}}^{\rm PC}]|\mathcal{B}_i\ra,   \nonumber\\
&A^{\rm{com}}(\mathcal{B}_i\to \mathcal{B}_f \overline{K}^0)=\frac{1}{f_K}\la \mathcal{B}_f|[U_+, H_{\rm{eff}}^{\rm PC}]|\mathcal{B}_i\ra,   \nonumber
\end{align}
where we have introduced the isospin $I$, $U$-spin and $V$-spin ladder operators with
\be \label{eq:IUV}
I_+|d\ra=|u\ra, \quad I_-|u\ra=|d\ra, \quad U_+|s\ra=|d\ra, \quad U_-|d\ra=|s\ra, \quad V_+|s\ra=|u\ra, \quad V_-|u\ra=|s\ra.  
\en
In Eq. (\ref{eq:commu}),  $\eta_8$ is the octet component of the $\eta$ and $\eta'$
\be
\eta=\cos\theta\eta_8-\sin\theta\eta_0, \qquad \eta'=\sin\theta\eta_8+\cos\theta\eta_0,
\en
with $\theta=-15.4^\circ$ \cite{Kroll}. For the decay constant $f_{\eta_8}$,  we shall follow \cite{Kroll} to use $f_{\eta_8}=f_8\cos\theta$ with $f_8=1.26 f_\pi$.
The hypercharge $Y$ is taken to be $Y=B+S-C$ \cite{Cheng:2018hwl}, where $B$, $C$ and $S$ are the quantum numbers of the baryon, charm and strangeness, respectively.

A straightforward calculation gives the following results:
\begin{align} \label{eq:Swave}
& A^{\rm{com}}(\Xi_{cc}^{++}\to\Xi_c^+ \pi^+)=\frac{1}{f_\pi}\left(
-a_{\Xi_c^+ \Xi_{cc}^+}
\right),   \qquad
A^{\rm{com}}(\Xi_{cc}^{++}\to\Xi_c^{'+} \pi^+)=\frac{1}{f_\pi}\left(
-a_{\Xi_c^{'+} \Xi_{cc}^+}
\right),\nonumber\\
&A^{\rm{com}}(\Xi_{cc}^{+}\to\Xi_c^+ \pi^0)=\frac{\sqrt{2}}{f_\pi}\left(  a_{\Xi_c^+ \Xi_{cc}^+}
\right), \qquad\quad~
A^{\rm{com}}(\Xi_{cc}^{+}\to\Xi_c^{'+} \pi^0)=\frac{\sqrt{2}}{f_\pi}\left(  a_{\Xi_c^{'+} \Xi_{cc}^+}
\right),  \nonumber \\
&A^{\rm{com}}(\Xi_{cc}^{+}\to\Xi_c^{+} \eta)=\frac{\sqrt{6}}{f_{\eta_8}} \left(a_{\Xi_c^+ \Xi_{cc}^+} \right),
\qquad\qquad
A^{\rm{com}}(\Xi_{cc}^{+}\to\Xi_c^{'+} \eta)=\frac{\sqrt{6}}{f_{\eta_8}} \left( a_{\Xi_c^{'+} \Xi_{cc}^+} \right), \nonumber\\
&A^{\rm{com}}(\Xi_{cc}^{+}\to\Sigma_c^{++} K^-)= \frac{2}{f_K} \left(a_{\Xi_c^{'+} \Xi_{cc}^+} \right), \qquad
A^{\rm{com}}(\Xi_{cc}^{+}\to\Lambda_c^+ \overline{K}^0)= \frac{1}{f_K}\left(a_{\Xi_c^+ \Xi_{cc}^+} \right),  \\
&A^{\rm{com}}(\Xi_{cc}^{+}\to\Sigma_c^{+} \overline{K}^0)= \frac{1}{f_K}\left(a_{\Xi_c^{'+} \Xi_{cc}^+} \right), \qquad\quad
A^{\rm{com}}(\Xi_{cc}^{+}\to\Xi_c^{0} \pi^+)=  \frac{1}{f_\pi}\left(a_{\Xi_c^+ \Xi_{cc}^+} \right), \nonumber\\
&A^{\rm{com}}(\Xi_{cc}^{+}\to\Xi_c^{'0} \pi^+)=  \frac{1}{f_\pi}\left(a_{\Xi_c^{'+} \Xi_{cc}^+} \right), \qquad\quad~
{{A^{\rm{com}}(\Xi_{cc}^{+}\to\Omega_c^0 K^+)=\frac{\sqrt{2}}{f_K} \left( a_{\Xi_c^{'+}\Xi_{cc}^+}
\right)
}}
,  \nonumber\\
&A^{\rm{com}}(\Omega_{cc}^{+}\to\Xi_c^+ \overline{K}^0)=  \frac{1}{f_K}\left(-a_{\Xi_c^+ \Xi_{cc}^+} \right),\qquad~
A^{\rm{com}}(\Omega_{cc}^{+}\to\Xi_c^{'+} \overline{K}^0)=  \frac{1}{f_K}\left(-a_{\Xi_c^{'+} \Xi_{cc}^+} \right),  \nonumber
\end{align}
where the baryonic matrix element $\la \mathcal{B}'|H_{\rm{eff}}^{\rm PC}|\mathcal{B}\ra$ is denoted by $a_{\mathcal{B}'\mathcal{B}}$. Evidently, all the $S$-wave amplitudes are governed by the matrix elements $a_{\Xi_c^+ \Xi_{cc}^+}$ and $a_{\Xi_c^{'+} \Xi_{cc}^+}$. We shall see shortly that this is also true for the  $P$-wave pole amplitudes.

\subsubsection{$P$-wave amplitudes}

We next turn to the nonfactorizable $P$-wave amplitudes given by
Eq. (\ref{eq:Bpole}). We have
\begin{align} \label{eq:Xicc++Pwave}
&B^{\rm{ca}}(\Xi_{cc}^{++}\to\Xi_c^+ \pi^+)=\frac{1}{f_\pi}\left(a_{\Xi_c^{+} \Xi_{cc}^{+}}\frac{m_{\Xi_{cc}^{++}}+m_{\Xi_{cc}^+}}{m_{\Xi_c^+}-m_{\Xi_{cc}^+}} g^{A(\pi^+)}_{\Xi_{cc}^+\Xi_{cc}^{++}}
\right), \nonumber\\
&B^{\rm{ca}}(\Xi_{cc}^{++}\to\Xi_c^{'+} \pi^+)=\frac{1}{f_\pi}\left(a_{\Xi_c^{'+} \Xi_{cc}^{+}}\frac{m_{\Xi_{cc}^{++}}+m_{\Xi_{cc}^+}}{m_{\Xi_c^{'+}}-m_{\Xi_{cc}^+}} g^{A(\pi^+)}_{\Xi_{cc}^+\Xi_{cc}^{++}}
\right),
\end{align}
for Cabibbo-favored  $\Xi_{cc}^{++}$  decays,
\be \label{eq:Xicc+Pwave}
B^{\rm{ca}}(\Xi_{cc}^{+}\to\Xi_c^+ \pi^0) &=& \frac{\sqrt{2}}{f_\pi}\Bigg(  a_{\Xi_c^+ \Xi_{cc}^+}\frac{2 m_{\Xi_{cc}^+}}{m_{\Xi_{c}^+}-m_{\Xi_{cc}^+}}
g^{A(\pi^0)}_{\Xi_{cc}^+\Xi_{cc}^+} +
g^{A(\pi^0)}_{\Xi_c^+ \Xi_c^+}
\frac{2m_{\Xi_c^+}}{m_{\Xi_{cc}^+}-m_{\Xi_c^+}} a_{\Xi_c^+ \Xi_{cc}^+}
\nonumber \\
&+& g^{A(\pi^0)}_{\Xi_c^+ \Xi_c^{'+}}
\frac{m_{\Xi_c^+}+m_{\Xi_c^{'+}}}{m_{\Xi_{cc}^+}-m_{\Xi_c^{'+}}} a_{\Xi_c^{'+} \Xi_{cc}^+}
\Bigg), \nonumber\\
B^{\rm{ca}}(\Xi_{cc}^{+}\to\Xi_c^{'+} \pi^0) &=& \frac{\sqrt{2}}{f_\pi}\Bigg(  a_{\Xi_c^{'+} \Xi_{cc}^+}\frac{2 m_{\Xi_{cc}^+}}{m_{\Xi_{c}^{'+}}-m_{\Xi_{cc}^+}}
g^{A(\pi^0)}_{\Xi_{cc}^+\Xi_{cc}^+}  +
g^{A(\pi^0)}_{\Xi_c^{'+} \Xi_c^+}
\frac{m_{\Xi_c^{'+}}+m_{\Xi_c^+}}{m_{\Xi_{cc}^+}-m_{\Xi_c^+}} a_{\Xi_c^+ \Xi_{cc}^+}
\nonumber \\
&+& g^{A(\pi^0)}_{\Xi_c^{'+} \Xi_c^{'+}}
\frac{2m_{\Xi_c^{'+}}}{m_{\Xi_{cc}^+}-m_{\Xi_c^{'+}}} a_{\Xi_c^{'+} \Xi_{cc}^+}
\Bigg), \nonumber\\
B^{\rm{ca}}(\Xi_{cc}^{+}\to\Xi_c^{+} \eta_8) &=& \frac{\sqrt{2}}{f_{\eta_8}}\Bigg(  a_{\Xi_c^+ \Xi_{cc}^+}\frac{2 m_{\Xi_{cc}^+}}{m_{\Xi_{c}^+}-m_{\Xi_{cc}^+}}
g^{A(\eta_8)}_{\Xi_{cc}^+\Xi_{cc}^+} +
g^{A(\eta_8)}_{\Xi_c^+ \Xi_c^+}
\frac{2m_{\Xi_c^+}}{m_{\Xi_{cc}^+}-m_{\Xi_c^+}} a_{\Xi_c^+ \Xi_{cc}^+}
\nonumber \\
&+& g^{A(\eta_8)}_{\Xi_c^+ \Xi_c^{'+}}
\frac{m_{\Xi_c^+}+m_{\Xi_c^{'+}}}{m_{\Xi_{cc}^+}-m_{\Xi_c^{'+}}} a_{\Xi_c^{'+} \Xi_{cc}^+}
\Bigg), \\
B^{\rm{ca}}(\Xi_{cc}^{+}\to\Xi_c^{'+} \eta_8) &=& \frac{\sqrt{2}}{f_{\eta_8}}\Bigg(  a_{\Xi_c^{'+} \Xi_{cc}^+}\frac{2 m_{\Xi_{cc}^+}}{m_{\Xi_{c}^{'+}}-m_{\Xi_{cc}^+}}
g^{A(\eta_8)}_{\Xi_{cc}^+\Xi_{cc}^+} +
g^{A(\eta_8)}_{\Xi_c^{'+} \Xi_c^+}
\frac{m_{\Xi_c^{'+}}+m_{\Xi_c^+}}{m_{\Xi_{cc}^+}-m_{\Xi_c^+}} a_{\Xi_c^+ \Xi_{cc}^+}
\nonumber \\
&+& g^{A(\eta_8)}_{\Xi_c^{'+} \Xi_c^{'+}}
\frac{2m_{\Xi_c^{'+}}}{m_{\Xi_{cc}^+}-m_{\Xi_c^{'+}}} a_{\Xi_c^{'+} \Xi_{cc}^+}
\Bigg), \nonumber\\
B^{\rm{ca}}(\Xi_{cc}^{+}\to\Sigma_c^{++} K^-) &=& \frac{1}{f_K}\left( g^{A(K^-)}_{\Sigma_{c}^{++}\Xi_c^+}
\frac{m_{\Sigma_c^{++}}+m_{\Xi_c^+}}{m_{\Xi_{cc}^+} -m_{\Xi_c^+}} a_{\Xi_c^+ \Xi_{cc}^+}
+
g^{A(K^-)}_{\Sigma_{c}^{++}\Xi_c^{'+}}
\frac{m_{\Sigma_c^{++}}+m_{\Xi_c^{'+}}}{m_{\Xi_{cc}^+} -m_{\Xi_c^{'+}}} a_{\Xi_c^{'+} \Xi_{cc}^+}
\right),  \nonumber \\
B^{\rm{ca}}(\Xi_{cc}^{+}\to\Lambda_c^+ \overline{K}^0) &=& \frac{1}{f_K}\left( g^{A(\overline{K}^0)}_{\Lambda_c^+ \Xi_c^+}
\frac{m_{\Lambda_c^+}+m_{\Xi_c^+}}{m_{\Xi_{cc}^+}-m_{\Xi_c^+}} a_{\Xi_c^+ \Xi_{cc}^+}
+
g^{A(\overline{K}^0)}_{\Lambda_c^+ \Xi_c^{'+}}
\frac{m_{\Lambda_c^+}+m_{\Xi_c^{'+}}}{m_{\Xi_{cc}^+}-m_{\Xi_c^{'+}}} a_{\Xi_c^{'+} \Xi_{cc}^+}
 \right), \nonumber \\
B^{\rm{ca}}(\Xi_{cc}^{+}\to\Sigma_c^{+} \overline{K}^0) &=&   \frac{1}{f_K}\left( g^{A(\overline{K}^0)}_{\Sigma_c^+ \Xi_c^+}
\frac{m_{\Sigma_c^+}+m_{\Xi_c^+}}{m_{\Xi_{cc}^+}-m_{\Xi_c^+}} a_{\Xi_c^+ \Xi_{cc}^+}
+
g^{A(\overline{K}^0)}_{\Sigma_c^+ \Xi_c^{'+}}
\frac{m_{\Sigma_c^+}+m_{\Xi_c^{'+}}}{m_{\Xi_{cc}^+}-m_{\Xi_c^{'+}}} a_{\Xi_c^{'+} \Xi_{cc}^+}
 \right), \nonumber \\
B^{\rm{ca}}(\Xi_{cc}^{+}\to\Xi_c^{0} \pi^+) &=&  \frac{1}{f_\pi}\left( g^{A(\pi^+)}_{\Xi_c^0 \Xi_c^+}
\frac{m_{\Xi_c^0}+m_{\Xi_c^+}}{m_{\Xi_{cc}^+}-m_{\Xi_c^+}} a_{\Xi_c^+ \Xi_{cc}^+}
+
g^{A(\pi^+)}_{\Xi_c^0 \Xi_c^{'+}}
\frac{m_{\Xi_c^0}+m_{\Xi_c^{'+}}}{m_{\Xi_{cc}^+}-m_{\Xi_c^{'+}}} a_{\Xi_c^{'+} \Xi_{cc}^+}
 \right),  \nonumber  \\
B^{\rm{ca}}(\Xi_{cc}^{+}\to\Xi_c^{'0} \pi^+) &=&  \frac{1}{f_\pi}\left( g^{A(\pi^+)}_{\Xi_c^{'0} \Xi_c^+}
\frac{m_{\Xi_c^{'0}}+m_{\Xi_c^+}}{m_{\Xi_{cc}^+}-m_{\Xi_c^+}} a_{\Xi_c^+ \Xi_{cc}^+}
+
g^{A(\pi^+)}_{\Xi_c^{'0} \Xi_c^{'+}}
\frac{m_{\Xi_c^{'0}}+m_{\Xi_c^{'+}}}{m_{\Xi_{cc}^+}-m_{\Xi_c^{'+}}} a_{\Xi_c^{'+} \Xi_{cc}^+}
 \right),  \nonumber \\
 B^{\rm{ca}}(\Xi_{cc}^{+}\to\Omega_c^0 K^+) &=& \frac{1}{f_K}\left( g^{A(K^+)}_{\Omega_c^0 \Xi_c^+}
\frac{m_{\Omega_c^0}+m_{\Xi_c^+}}{m_{\Xi_{cc}^+}-m_{\Xi_c^+}} a_{\Xi_c^+ \Xi_{cc}^+}
+
g^{A(K^+)}_{\Omega_c^0 \Xi_c^{'+}}
\frac{m_{\Omega_c^0}+m_{\Xi_c^{'+}}}{m_{\Xi_{cc}^+}-m_{\Xi_c^{'+}}} a_{\Xi_c^{'+} \Xi_{cc}^+}
\right),
\nonumber
\en
for Cabibbo-favored $\Xi_{cc}^{+}$  decays, and
\begin{align} \label{eq:OmegaccPwave}
&B^{\rm{ca}}(\Omega_{cc}^{+}\to\Xi_c^+ \overline{K}^0)= \frac{1}{f_K}\left(
a_{\Xi_c^+ \Xi_{cc}^+} \frac{m_{\Omega_{cc}^+}+m_{\Xi_{cc}^+}}{m_{\Xi_c^+}-m_{\Xi_{cc}^+}}g^{A(\overline{K}^0)}_{\Xi_{cc}^+ \Omega_{cc}^+}\right), \nonumber  \\
&B^{\rm{ca}}(\Omega_{cc}^{+}\to\Xi_c^{'+} \overline{K}^0)=
 \frac{1}{f_K}\left(
a_{\Xi_c^{'+} \Xi_{cc}^+} \frac{m_{\Omega_{cc}^+}+m_{\Xi_{cc}^+}}{m_{\Xi_c^{'+}}-m_{\Xi_{cc}^+}}g^{A(\overline{K}^0)}_{\Xi_{cc}^+ \Omega_{cc}^+}\right),
\end{align}
for Cabibbo-favored $\Omega_{cc}^{+}$  decays.

\subsection{Hadronic matrix elements and axial-vector form factors}
There are two types of non-perturbative quantities involved in the nonfactorizable amplitudes: hadronic matrix elements and axial-vector form factors. We will calculate them within the
framework of the MIT bag model \cite{MIT}.

\subsubsection{Hadronic matrix elements}
The baryonic matrix element $a_{\B'\B}$ plays an important role in both $S$-wave
and $P$-wave amplitudes. Its general expression in terms of the effective Hamiltonian Eq. (\ref{eq:effH}) is given by
\begin{align}
&a_{\B'\B} \equiv  \la \B'|\mathcal{H}_{\rm{eff}}^{\rm{PC}}|\B\ra
=\frac{G_F}{2\sqrt{2}} V_{cs} V^*_{ud} c_-\la \B' |O_- |\B\ra,
\end{align}
where $O_\pm=(\bar{s}c)(\bar{u}d)\pm(\bar{s}d)(\bar{u}c)$ and $c_\pm=c_1\pm c_2$.
The matrix element of $O_+$ vanishes as this operator is symmetric in color indices. In the MIT bag model, the matrix elements $a_{\Xi_c^+ \Xi_{cc}^+}$ and $a_{\Xi_c^{'+} \Xi_{cc}^+}$
are given by \footnote{For the evaluation of baryon matrix elements and form factors in the MIT bag model, see e.g. \cite{Cheng:1991sn,Cheng:1993gf}.
}
\begin{align}
&\la\Xi_c^+ | O_-|\Xi_{cc}^+\ra
=4\sqrt{6} X_2  (4\pi),  \qquad\qquad
\la\Xi^{'+}_c | O_-|\Xi_{cc}^+\ra
=-\frac{4\sqrt{2}}{3} X_1  (4\pi),
\end{align}
where we have introduced the bag integrals $X_1$ and $X_2$
\begin{align} \label{eq:X}
&X_1=\int^R_0 r^2 dr (u_s v_u-v_s u_u)(u_c v_d -v_c u_d)=3.56\times 10^{-6}, \nonumber\\
&X_2=\int^R_0 r^2 dr (u_s u_u+v_s v_u)(u_c u_d +v_c v_d)=1.74\times 10^{-4}.
\end{align}
To obtain numerical results, we have employed the following bag parameters
\be
m_u=m_d=0, \quad m_s=0.279~{\rm GeV}, \quad m_c=1.551~{\rm GeV}, \quad R=5~{\rm GeV}^{-1},
\en
where $R$ is the radius of the bag.

\subsubsection{Axial-vector form factors}
The axial-vector form factor in the static limit can be expressed in the bag model as
\begin{equation}
g^{A(P)}_{\mathcal{B}'\mathcal{B}}=\la\mathcal{B}'\uparrow | b_{q_1}^\dagger b_{q_2}\sigma_z|
\mathcal{B}\uparrow\ra \int d^3\bm{r}\left(u_{q_1}u_{q_2}-\frac13 v_{q_1}v_{q_2}\right),
\end{equation}
where $\sigma_z$ is the $z$-component of Pauli matrices. 
The relevant results are
\be \label{eq:gA}
&& g^{A(\pi^+)}_{\Xi_{cc}^+\Xi_{cc}^{++}}=-\frac13 (4\pi Z_1),
\qquad
g^{A(\pi^0)}_{\Xi_{cc}^+\Xi_{cc}^+}=   \frac16 (4\pi Z_1),
\qquad\quad~
g^{A(\eta_8)}_{\Xi_{cc}^+\Xi_{cc}^+}=   -\frac{1}{6\sqrt{3}}(4\pi Z_1),
\nonumber\\
&& g^{A(K^-)}_{\Sigma_{c}^{++}\Xi_{c}^+}=   \frac{\sqrt{6}}{3}(4\pi Z_2),\qquad
g^{A(K^-)}_{\Sigma_{c}^{++}\Xi_{c}^{'+}}=   \frac{2\sqrt{2}}{3}(4\pi Z_2),
\quad~
g^{A(\overline{K}^0)}_{\Xi_{cc}^+\Omega_{cc}^+}=   -\frac13(4\pi Z_2),\nonumber\\
&& g^{A(\overline{K}^0)}_{\Lambda_{c}^+\Xi_{c}^{'+}}=   -\frac{\sqrt{3}}{3}(4\pi Z_2), \quad~~
g^{A(\overline{K}^0)}_{\Sigma_{c}^+\Xi_{c}^+}=   \frac{\sqrt{3}}{3}(4\pi Z_2), \qquad~
g^{A(\overline{K}^0)}_{\Sigma_{c}^+\Xi_{c}^{'+}}=   \frac23(4\pi Z_2),
\nonumber \\
&& g^{A(\pi^+)}_{\Xi_{c}^{0}\Xi_{c}^{'+}}=   -\frac{\sqrt{3}}{3}(4\pi Z_1), \qquad
g^{A(\pi^+)}_{\Xi_{c}^{'0}\Xi_{c}^+}=   -\frac{\sqrt{3}}{3}(4\pi Z_1), \quad~~
g^{A(\pi^+)}_{\Xi_{c}^{'0}\Xi_{c}^{'+}}=   \frac23 (4\pi Z_1),
\\
&& g^{A(\pi^0)}_{\Xi_c^{'+}\Xi_c^+}  = -\frac{\sqrt{3}}{6}(4\pi Z_1) ,\qquad
g^{A(\pi^0)}_{\Xi_c^{'+}\Xi_c^+} = -\frac{\sqrt{3}}{6}(4\pi Z_1), \quad~~
g^{A(\pi^0)}_{\Xi_c^{'+}\Xi_c^{'+}}=\frac13 (4\pi Z_1),
\nonumber\\
&& g^{A(\eta_8)}_{\Xi_c^{'+}\Xi_c^+}  = -\frac{1}{2}(4\pi Z_1) ,\qquad~~
g^{A(\eta_8)}_{\Xi_c^{'+}\Xi_c^+} = -\frac{1}{2}(4\pi Z_1),\qquad~~
g^{A(\eta_8)}_{\Xi_c^{'+}\Xi_c^{'+}}=-\frac{\sqrt{3}}{9} (4\pi Z_1),\nonumber\\
&& g_{\Omega_c^0 \Xi_c^+}^{A(K^+)}=-\frac{\sqrt{6}}{3}(4\pi Z_2),\qquad
g_{\Omega_c^0 \Xi_c^{'+}}^{A(K^+)}=\frac{2\sqrt{2}}{3}(4\pi Z_2),
\nonumber 
\en 
and
\be
g^{A(\overline{K}^0)}_{\Lambda_{c}^+\Xi_{c}^+}=   0, \qquad\qquad g^{A(\pi^+)}_{\Xi_{c}^{0}\Xi_{c}^+}=   0,   \qquad\qquad
g^{A(\pi^0)}_{\Xi_c^+ \Xi_c^+}=0,\qquad\qquad
  g^{A(\eta_8)}_{\Xi_c^+ \Xi_c^+}=0,  
\en
where the auxiliary bag integrals are given by
\begin{equation}
Z_1=\int r^2 dr\left(u_u^2 -\frac13 v_u^2\right),\qquad
Z_2=\int r^2 dr \left(u_u u_s -\frac13 v_u v_s\right).
\end{equation}
Numerically,
$(4\pi) Z_1= 0.65$ and  $(4\pi)Z_2=0.71$.

\section{Results and discussions}
\label{sec:num}

\subsection{Numerical results and discussions}

For numerical calculations, we shall use the Wilson coefficients $c_1(\mu)=1.346$ and $c_2(\mu)=-0.636$ evaluated at the scale $\mu=1.25$ GeV with $\Lambda^{(4)}_{\overline {\rm MS}}=325$ MeV \cite{Buchalla}. We follow \cite{Cheng:2018hwl} to use the Wilson coefficients $a_1=1.26\pm0.02$ and $a_2=-0.45\pm0.05$, corresponding to $N_c^{\rm eff}\approx 7$. Recall that the value of $|a_2|$ is determined from the measurement of $\Lambda_c^+\to p\phi$ \cite{BES:pphi}, which proceeds only through the internal $W$-emission diagram. For the CKM matrix elements we use $V_{ud}=0.9743$ and $V_{cs}=0.9735$.  The mass of the
$\Omega_{cc}^+$ is taken to be  $3.712\,{\rm{GeV}}$ from lattice QCD \cite{LQCD:Omegacc}. For the $\Xi_{cc}^+$, we assume that it has the same mass as the $\Xi_{cc}^{++}$ which is taken to be 3621 MeV from Eq. (\ref{eq:Xiccmass}). This is justified because the isospin splitting in the doubly charmed baryons with light quarks has been estimated to be very small, $m_{\Xi_{cc}^{++}}-m_{\Xi_{cc}^+}={\cal O}(1.5)$ MeV (see \cite{Karliner:2017gml} and references therein).

To calculate branching fractions we need to know the lifetimes of the doubly charmed baryons $\Xi_{cc}^+$ and $\Omega_{cc}^+$ in addition to the lifetime of $\Xi_{cc}^{++}$ measured by the LHCb. The lifetimes of doubly charmed hadrons have been analyzed within the framework of heavy quark expansion \cite{Kiselev:1999,Kiselev:2002,Guberina,Chang,Karliner:2014,Cheng:doubly,Berezhnoy}.
Lifetime differences arise from spectator effects such as $W$-exchange and Pauli interference.  The $\Xi_{cc}^{++}$ baryon is longest-lived in the doubly charmed baryon system owing to the destructive Pauli interference absent in the $\Xi_{cc}^+$ and $\Omega_{cc}^+$.
As shown in \cite{Cheng:doubly}, it is necessary to take into account dimension-7 spectator effects in order to obtain  the $\Xi_{cc}^{++}$ lifetime consistent with the LHCb measurement (see Eq. (\ref{eq:lifetime})). It is difficult to make a precise quantitative statement on the lifetime of $\Omega_{cc}^+$ because of the uncertainties associated with the dimension-7 spectator effects in the $\Omega_{cc}^+$. It was estimated in \cite{Cheng:doubly} that $\tau(\Omega_{cc}^+)$ lies in the range of $(0.75\sim 1.80)\times 10^{-13}s$. For our purpose, we shall take the mean lifetime $\tau(\Omega_{cc}^+)=1.28\times 10^{-13}s$. On the contrary, the lifetime of $\Xi_{cc}^+$ is rather insensitive to the variation of dimension-7 effects and $\tau(\Xi_{cc}^+)=0.45\times 10^{-13}s$ was obtained \cite{Cheng:doubly}.
The lifetimes of doubly charmed baryons respect the hierarchy pattern  $\tau(\Xi_{cc}^{++})>\tau(\Omega_{cc}^+)>\tau(\Xi_{cc}^+)$.

\begin{table}[t]
\caption{The predicted $S$- and $P$-wave amplitudes of Cabibbo-favored $\B_{cc}\to{\cal B}_c+P$ decays in units of $10^{-2} G_F{\rm GeV}^2$.
 Branching fractions (in units of $10^{-2}$) and the decay asymmetry parameter $\alpha$ are shown in the last two columns. For lifetimes we use $\tau(\Xi_{cc}^{++})=2.56\times10^{-13}s$, $\tau(\Xi_{cc}^{+})=0.45\times10^{-13}s$ and $\tau(\Omega_{cc}^+)=1.28\times10^{-13}s$ (see the main text).
 }   \label{tab:BF}
 \vspace{0.2cm}
\centering
\begin{ruledtabular}
\begin{tabular}{ l rrrrrr|cr}
 Channel & $A^{\rm{fac}}$ &  $A^{\rm{com}}$ & $A^{\rm{tot}}$ & $B^{\rm{fac}}$ &  $B^{\rm{ca}}$ & $B^{\rm{tot}}$~~ & $\mathcal{B}_{\rm{theo}}$
 &  $\alpha_{\rm{theo}}$\\
 \hline
   \midrule
$\Xi_{cc}^{++} \to \Xi^{+}_{c}\pi^+$  & $7.40$  & $-10.79$  &$-3.38$    & $-15.06$  &   $18.91$   & $3.85$~~  & $0.69$   & $-0.41$     \\
$\Xi_{cc}^{++} \to \Xi^{'+}_{c}\pi^+$ &  $4.49$  & $-0.04$  &$4.45$    & $-48.50$  &   $0.06$   &
$-48.44$~~  & $4.65$   & $-0.84$     \\
$\Xi_{cc}^{++} \to \Sigma _c^{++}\overline{K}^0$ & $-2.67$  & $0$  &$-2.67$    & $25.11$  &   $0$   &
$25.11$~~  & $1.36$   & $-0.89$     \\
\hline
$\Xi_{cc}^{+} \to \Xi_{c}^{0}\pi^+$ & $8.52$  & $10.79$  &$19.31$    & $-16.46$  &   $-0.08$   &
$-16.54$~~ & 3.84   & $-0.31$   \\
$\Xi_{cc}^{+} \to \Xi_{c}^{'0}\pi^+$& $5.05$  & $0.04$  &$5.09$    & $-52.31$  &   $-17.63$  &
$-69.94$~~  &  1.55   & $-0.73$   \\
$\Xi_{cc}^{+} \to \Xi^{+}_{c}\pi^0$ &  $0$  & $15.26$  &$15.26$    & $0$  &   $-10.49$ &
 $-10.49$~~  & 2.38& $-0.25$     \\
$\Xi_{cc}^{+} \to \Xi^{'+}_{c}\pi^0$ &  $0$  & $0.06$  &$0.06$    & $0$  &   $-24.97$  &
 $-24.97$~~  & 0.17   &  $-0.03$     \\
$\Xi_{cc}^{+} \to \Xi_{c}^{+} \eta$& $0$  & $21.75$  & $21.75$  & $0$  &   $4.86$   &
  $4.86$~~  & $4.18$  & $0.07$   \\
$\Xi_{cc}^{+} \to \Xi_{c}^{'+} \eta$& $0$  & $0.09$   & $0.09$   & 0
 &  $-17.87$ &  $-17.87$~~
 &  $0.05$  & $-0.07$   \\
$\Xi_{cc}^{+} \to \Sigma _c^{++}K^{-}$ & $0$  & $0.07$  &$0.07$    & $0$  &   $22.14$   &
$22.14$~~  & $0.13$  & $0.04$     \\
$\Xi_{cc}^{+} \to \Lambda_c^+\overline{K}^0$ &  $-3.37$  & $8.90$  &$5.53$    & $5.62$  &   $-0.07$   &
$5.55$~~  & $0.31$   & $0.40$     \\
$\Xi_{cc}^{+} \to \Sigma^+_c\overline{K}^0$ & $-2.17$  & $0.04$  &$-2.14$    & $19.37$  &   $15.64$   &
$35.02$~~  & $0.38$  & $-0.62$   \\
$\Xi_{cc}^{+} \to \Omega_c^0 K^+$  & $0$ & $0.05$  & $0.05$
&  $0$ & $-22.98$   &  $-22.98$~~
& $0.06$ & $-0.03$
 \\
\hline
$\Omega_{cc}^+ \to \Omega^0_c\pi^+$  &  $5.71$  & $0$  &$5.71$    & $-67.48$  &   $0$   &
$-67.48$~~  & $3.96$  & $-0.83$     \\
$\Omega_{cc}^+ \to \Xi^{+}_{c}\overline{K}^0$ &  $2.62$  & $-8.90$  &$-6.28$    & $-5.29$  &   $13.40$  &
$8.11$~~  & $1.15$   & $-0.45$     \\
$\Omega_{cc}^+ \to \Xi^{'+}_{c}\overline{K}^0$ &  $-1.68$  & $-0.04$  &$-1.72$    & $17.44$  &   $0.06$   &
$17.50$~~  & $0.29$  & $-0.88$     \\
\end{tabular}
\end{ruledtabular}
\end{table}

Factorizable and nonfactorizable amplitudes, branching fractions and decay asymmetries for Cabibbo-favored two-body decays $\B_{cc}\to \mathcal{B}_c P$ calculated in this work are summarized in Table \ref{tab:BF}.
The channel $\Xi_{cc}^{++}\to \Xi_c^+\pi^+$ is the first two-body decay mode observed by the LHCb in the doubly charmed baryon sector. However, our prediction of $0.69\%$\footnote{A straightforward calculation in our framework yields a branching fraction of $0.66\%$ and $\alpha=0.04$ for $\Xi_{cc}^{++}\to \Xi_c^+\pi^+$. The tiny decay asymmetry is due to a large cancellation between $B^{\rm fac}(=-15.06)$ and $B^{\rm ca}(=14.69)$. Normally, a huge cancellation between two terms will lead to a unreliable prediction. Hence, we have replaced $B^{\rm ca}$ by $B^{\rm pole}(=18.91)$ and used $g_{\Xi_{cc}^{++}\Xi_{cc}^+\pi^+}=-15.31$ \cite{Sharma:2017txj}, where the sign of the strong coupling is fixed by the axial-vector form factor $g^{A(\pi^+)}_{\Xi_{cc}^+\Xi_{cc}^{++}}$ given in Eq. (\ref{eq:gA}).
}
for its branching fraction is substantially smaller than the results of $(3\sim 9)\%$ given in the literature (see Table \ref{tab:comparison2} below). This is ascribed to the destructive interference between factorizable and nonfactorizable contributions for both $S$- and $P$-wave amplitudes (see Table \ref{tab:BF}).
If we turn off the nonfactorizable terms, we will have a branching fraction of order $3.6\%$.
In the literature, nonfactorizable effects in $\Xi_{cc}^{++}\to \Xi_c^+\pi^+$  have been considered in \cite{Gutsche:2018msz} and partially in \cite{Sharma:2017txj} (c.f. Table \ref{tab:Xicpi}).
It is very interesting to notice that our calculation agrees with \cite{Gutsche:2018msz}
even though the estimation of nonfactorizable effects  is based on entirely different approaches: current algebra and the pole model in this work and the covariant confined quark model in \cite{Gutsche:2018msz}.  On the contrary, a large constructive interference
in the $P$-wave amplitude was found in \cite{Sharma:2017txj},\footnote{The pole amplitudes obtained by Dhir and Sharma  shown in the tables of  \cite{Sharma:2017txj,Dhir:2018twm} were calculated using their Eq. (8) without a minus sign in front of $\sum_n$. Therefore, it is necessary to assign an extra minus sign in order to get $B^{\rm pole}$. For example, $B^{\rm pole}(\Xi_{cc}^{++}\to\Xi_c^+\pi^+)$ should read $-0.372$ rather than 0.372 for the flavor independent case
(see Table III of \cite{Sharma:2017txj}). Hence, the pole and factorizable $P$-wave amplitudes in  $\Xi_{cc}^{++}\to\Xi_c^+\pi^+$ interfere constructively in \cite{Sharma:2017txj}.
}
while nonfactorizable corrections to the $S$-wave one were not considered. This leads to a branching fraction of order $(7-9)\%$ ($(13-16)\%$) for flavor-independent (flavor-dependent) pole amplitudes.

\begin{table}[t!]
\caption{Comparison of the predicted $S$- and $P$-wave amplitudes (in units of $10^{-2} G_F{\rm GeV}^2$) of some Cabibbo-favored decays $\B_{cc}\to\B_c+P$ decays  in various approaches.
Branching fractions (in unit of $10^{-2}$) and the decay asymmetry parameter $\alpha$ are shown in the last two columns. We have converted the helicity amplitudes in Gutsche {\it et al.} \cite{Gutsche:2018msz} into the partial-wave ones. For the predictions of Dhir and Sharma \cite{Sharma:2017txj,Dhir:2018twm}, we quote the flavor-independent pole amplitudes  and two different models for $\B_{cc}\to\B_c$ transition form factors: nonrelativistic quark  model (abbreviated as N) and heavy quark effective theory (H). All the model results have been normalized using the lifetimes
$\tau(\Xi_{cc}^{++})=2.56\times10^{-13}s$, $\tau(\Xi_{cc}^{+})=0.45\times10^{-13}s$ and $\tau(\Omega_{cc}^+)=1.28\times10^{-13}s$.
 }   \label{tab:Xicpi}
 \vspace{0.2cm}
\centering
\begin{ruledtabular}
\begin{tabular}{ l rrrrrr|cr}
& $A^{\rm{fac}}$ &  $A^{\rm{nf}}$ & $A^{\rm{tot}}$ & $B^{\rm{fac}}$ &  $B^{\rm{nf}}$ & $B^{\rm{tot}}$~~ & $\mathcal{B}_{\rm{theo}}$
 &  $\alpha_{\rm{theo}}$\\
 \hline
   \midrule
 $\Xi_{cc}^{++} \to \Xi^{+}_{c}\pi^+$  &  & &  & & & &   &  \\
This work  & $7.40$  & $-10.79$  &$-3.38$    & $-15.06$  &   $18.91$   &
$3.85$~~  & $0.69$   & $-0.41$     \\
Gutsche {\it et al.} & $-8.13$  & $11.50$  & $3.37$    & $12.97$  &   $-18.53$   &
$-5.56$~~  & $0.71$   & $-0.57$     \\
Dhir $\&$ Sharma (N)&  $7.38$  &  0 & $7.38$    & $-16.77$  &   $-24.95$   &
$-41.72$~~  & $6.64$   & $-0.99$     \\
\qquad (H) &  $9.52$  & 0  & $9.52$    & $-19.45$  &   $-24.95$   &
$-44.40$~~  & $9.19$   & $-0.99$     \\
\hline
 $\Xi_{cc}^{++} \to \Xi^{'+}_{c}\pi^+$  &  & &  & & & &   &  \\
This work &  $4.49$  & $-0.04$  &$4.45$    & $-48.50$  &   $0.06$   &
$-48.44$~~  & $4.65$   & $-0.84$     \\
Gutsche {\it et al.} &  $-4.34$  & $-0.11$  & $-4.45$    & $37.59$  &   $1.37$   &
$38.96$~~  & $3.39$   & $-0.93$     \\
Dhir $\&$ Sharma (N) &  $4.29$  & $0$  &$4.29$    & $-53.65$  &   $0$   &
$-53.65$~~  & $5.39$   & $-0.78$     \\
\qquad (H) &  $5.10$  & $0$  & $5.10$    & $-62.37$  &   $0$   &
$-62.37$~~  & $7.34$   & $-0.79$     \\
\hline
 $\Xi_{cc}^{+} \to \Xi^{0}_{c}\pi^+$  &  & &  & & & &   &  \\
This work & $8.52$  & $10.79$  &$19.31$    & $-16.46$  &   $-0.08$   &
$-16.54$~~ & 3.84   & $-0.31$      \\
Dhir $\&$ Sharma (N) &  $7.38$  & $0$  &$7.38$    & $-16.77$  &   $28.30$   &
$11.54$~~  & $0.59$   & $0.54$     \\
\qquad (H) &  $9.59$  & $0$  & $9.59$    & $-19.45$  &   $28.30$   &
$8.85$~~  & $0.95$   & $0.34$     \\
\hline
$\Omega_{cc}^+ \to \Xi^{+}_{c}\overline{K}^0$ &  & &  & & & &   &  \\
This work &  $2.62$  & $-8.90$  &$-6.28$    & $-5.29$  &   $13.40$  &
$8.11$~~  & $1.15$   & $-0.45$     \\
Gutsche {\it et al.} & $-4.02$  & $12.17$  & $8.15$    & $6.20$  &   $-19.23$   &
$-13.02$~~  & $1.98$   & $-0.54$     \\
Dhir $\&$ Sharma (N) &  $3.42$  & $0$  &$3.42$    & $-8.12$  &   $-25.22$   &
$-33.33$~~  & $1.36$   & $-0.85$     \\
\qquad (H) &  $5.57$  & $0$  & $5.57$    & $-11.54$  &   $-25.22$   &
$-36.75$~~  & $2.12$   & $-0.98$     \\
\hline
$\Omega_{cc}^+ \to \Xi^{'+}_{c}\overline{K}^0$ &  & &  & & & &   &  \\
This work &  $-1.68$  & $-0.04$  &$-1.72$    & $17.44$  &   $0.06$   &
$17.50$~~  & $0.29$  & $-0.88$     \\
Gutsche {\it et al.} & $2.26$  & $-0.11$  & $2.14$    & $-17.34$  &   $0.69$   &
$-16.64$~~  & $0.31$   & $-0.97$     \\
Dhir $\&$ Sharma (N) &  $-2.15$  & $0$  &$-2.15$    & $26.8$  &   $0$   &
$26.8$~~  & $0.61$   & $-0.79$     \\
\qquad (H) & $-2.95$  & $0$  &$-2.95$    & $37.6$  &   $0$   &
$37.6$~~  & $1.19$   & $-0.78$     \\
\end{tabular}
\end{ruledtabular}
\end{table}

Since the absolute branching fractions of $(\Lambda_c^+,\Xi_c^+)\to p K^-\pi^+$ have been measured with the results $\B(\Lambda_c^+\to p K^-\pi^+)=(6.28\pm0.32)\%$ \cite{Tanabashi:2018oca} and $\B(\Xi_c^+\to p K^-\pi^+)=(0.45\pm0.21\pm0.07)\%$ \cite{Li:2019atu}, it follows from Eq. (\ref{eq:Xicpi}) that
\begin{equation}
{\mathcal{B}(\Xi_{cc}^{++}\to\Xi_c^+\pi^+) \over
\mathcal{B}(\Xi_{cc}^{++}\to\Lambda_c^+ K^- \pi^+\pi^+)}=0.49\pm0.27\,,
\end{equation}
where the uncertainty is dominated by the decay rate of $\Xi_c^+$ into $p K^-\pi^+$. Although the rate of $\Xi_{cc}^{++}\to\Lambda_c^+ K^- \pi^+\pi^+$ is unknown, it is plausible to assume that $\B(\Xi_{cc}^{++}\to\Lambda_c^+ K^- \pi^+\pi^+)\approx {2\over 3}\B(
\Xi_{cc}^{++}\to\Sigma_c^{++}\overline{K}^{*0})$. Since $\Xi_{cc}^{++}\to\Sigma_c^{++}\overline{K}^{*0}$ is a purely factorizable process, its rate can be reliably estimated once the relevant form factors are determined. Taking the latest prediction $\B(\Xi_{cc}^{++}\to\Sigma_c^{++}\overline{K}^{*0})=5.61\%$ from \cite{Gutsche:2019iac} as an example,
\footnote{The branching fraction is given by $(5.40^{+5.59}_{-3.66})\%$ in the approach of final-state rescattering \cite{Jiang:2018oak}.}
we obtain
\be
\mathcal{B}(\Xi_{cc}^{++}\to\Xi_c^+\pi^+)_{\rm expt}\approx (1.83\pm1.01)\%.
\en
Therefore, our prediction of $\mathcal{B}(\Xi_{cc}^{++}\to\Xi_c^+\pi^+)\approx 0.7\%$ is consistent with the experimental value but in the lower end. In future study, it is important to pin down the branching fraction of this mode both experimentally and theoretically.

In contrast to $\Xi_{cc}^{++}\to\Xi_c^+\pi^+$, we find a large constructive interference between factorizable and nonfactorizable $S$-wave amplitudes in $\Xi_{cc}^+\to \Xi_c^0\pi^+$, whereas Dhir and Sharma \cite{Sharma:2017txj} obtained a large destructive interference in $P$-wave amplitudes  (see Table \ref{tab:Xicpi}). Hence, the predicted rate of $\Xi_{cc}^+\to \Xi_c^0\pi^+$ in \cite{Sharma:2017txj} is rather suppressed compared to ours.
The hierarchy pattern ${\cal B}(\Xi_{cc}^{+}\to \Xi_c^0\pi^+)\gg {\cal B}(\Xi_{cc}^{++}\to \Xi_c^+\pi^+)$ is the analog of ${\cal B}(\Xi_{c}^{0}\to \Xi^-\pi^+)\gg {\cal B}(\Xi_{c}^{+}\to \Xi^0\pi^+)$ we found in \cite{Zou:2019kzq}. It should be noticed that
the hierarchy pattern ${\cal B}(\Xi_{cc}^{+}\to \Xi_c^0\pi^+)\ll {\cal B}(\Xi_{cc}^{++}\to \Xi_c^+\pi^+)$ obtained in \cite{Sharma:2017txj} is opposite to ours.

The large branching fraction of order 3.8\% for $\Xi_{cc}^{+}\to\Xi_c^0\pi^+$ may enable experimentalists to search for the $\Xi_{cc}^+$ through this mode.
That is, $\Xi_{cc}^+$ is reconstructed through the $\Xi_{cc}^+\to\Xi_c^0\pi^+$ followed by the decay chain $\Xi_c^0\to \Xi^-\pi^+\to p\pi^-\pi^-\pi^+$. Another popular way for the search of $\Xi_{cc}^+$ is through the processes $\Xi_{cc}^{+}\to \Lambda_c^+ K^-\pi^+$ and $\Lambda_c^+\to pK^-\pi^+$ \cite{Aaij:Xicc+,Yu:2019lxw}. \footnote{An estimate of the branching fraction of $\Xi_{cc}^{+}\to \Lambda_c^+ K^-\pi^+$ can be made by assuming $\B(\Xi_{cc}^{+}\to \Lambda_c^+ K^-\pi^+)\approx \B(\Xi_{cc}^{+}\to \Sigma^{++}K^-) +{2\over 3}\B(\Xi_{cc}^{+}\to \Lambda_c^+\overline{K}^{*0})$. Since $\B(\Xi_{cc}^{+}\to \Sigma^{++}K^-)\approx 0.13\%$ in our work, while $\B(\Xi_{cc}^{+}\to \Lambda_c^+\overline{K}^{*0})=(0.48^{+0.53}_{-0.33})\%$ is obtained in the final-state rescattering approach \cite{RHLi,Jiang:2018oak} for $\tau(\Xi_{cc}^+)=0.45\times 10^{-13}s$, it appears that $\B(\Xi_{cc}^{+}\to \Lambda_c^+ K^-\pi^+)$ is not more than 0.8\%.
}

\begin{table}[!]
\begin{center}
\footnotesize{
\caption{Predicted branching fractions (in $\%$) of Cabibbo-favored doubly charmed baryon decays by different groups.
For the predictions of Dhir and Sharma \cite{Sharma:2017txj,Dhir:2018twm}, we quote the flavor-independent pole amplitudes and two different models for $\B_{cc}\to\B_c$ transition form factors: nonrelativistic quark  model (abbreviated as N) and heavy quark effective theory (H).
For the results of Gutsche {\it et al.} \cite{Gutsche:2018msz,Gutsche:2017hux,Gutsche:2019iac}, we quote the latest ones from \cite{Gutsche:2019iac}.
All the model results have been normalized using the lifetimes
$\tau(\Xi_{cc}^{++})=2.56\times10^{-13}s$, $\tau(\Xi_{cc}^{+})=0.45\times10^{-13}s$ and $\tau(\Omega_{cc}^+)=1.28\times10^{-13}s$.}\label{tab:comparison2}
 \vspace{0.3cm}
\begin{ruledtabular}
\begin{tabular}{ l c c c c c c c  }
  Mode &  Our  & Dhir  & Gutsche \textit{et al.} & Wang  & Gerasimov & Ke & Shi\\
  & & \textit{et al.} \cite{Sharma:2017txj,Dhir:2018twm} &
   \cite{Gutsche:2018msz,Gutsche:2017hux,Gutsche:2019iac} & \textit{et al.} \cite{Wang:2017mqp}   & \textit{et al.} \cite{Gerasimov:2019jwp} & \textit{et al.} \cite{Ke:2019lcf} & \textit{et al.} \cite{Shi:2019hbf} \\
\midrule\hline
$\Xi_{cc}^{++}\to \Xi_c^{+}\pi^+$  & 0.69 & 6.64 (N) & 0.70 & 6.18  & 7.01 & $3.48\pm0.46$ & $3.1\pm0.4$ \\
&  & 9.19 (H) & & & &  \\
$\Xi_{cc}^{++}\to \Xi_c^{'+}\pi^+$  & 4.65 & 5.39 (N) & 3.03 & 4.33  & 5.85 & $1.96\pm0.24$ & $0.93\pm0.19$\\
&  & 7.34 (H) & & &  \\
$\Xi_{cc}^{++}\to \Sigma_c^{++}\overline{K}^0$ & 1.36 & 2.39 (N) &  $1.25$ & \\
&  & 4.69 (H) & & &  \\
\hline
$\Xi_{cc}^{+} \to \Xi_{c}^{0}\pi^+$ &  $3.84$   & 0.59 (N) & & 1.08   & 1.23 & $0.61\pm0.08$ & $0.53\pm0.08$ \\
&  & 0.95 (H) & & & &  \\
$\Xi_{cc}^{+} \to \Xi_{c}^{'0}\pi^+$  & $1.55$   &  1.49 (N) &  & 0.76  & 1.04 & $0.35\pm0.04$ & $0.16\pm0.03$ \\
&  & 2.12 (H) & & & &  \\
$\Xi_{cc}^{+} \to \Lambda_c^+\overline{K}^0$  & $0.31$ & 0.27 (N) &      \\
&  & 0.37 (H) & & & &  \\
$\Xi_{cc}^{+} \to \Sigma^+_c\overline{K}^0$  & $0.38$  & 0.59 (N)   \\
&  & 0.90 (H) & & & &  \\
$\Xi_{cc}^{+} \to \Xi^{+}_{c}\pi^0$  & $2.38$  &  0.50     \\
$\Xi_{cc}^{+} \to \Xi^{'+}_{c}\pi^0$   & $0.17$   & 0.054     \\
$\Xi_{cc}^{+} \to \Xi_{c}^{+} \eta$& $4.18$  &  0.063  &    \\
$\Xi_{cc}^{+} \to \Xi_{c}^{'+} \eta$& $0.05$  &  0.036  &    \\
$\Xi_{cc}^{+} \to \Sigma _c^{++}K^{-}$  & $0.13$  &  0.22    \\
$\Xi_{cc}^{+} \to \Omega_c^0 K^+$  & 0.06 & 0.10 & & & & \\
\hline
$\Omega_{cc}^+ \to \Omega^{0}_{c}\pi^+$ & 3.96 & 5.38 (N) & 3.08 & 3.34   & 5.30 & & $0.55\pm0.23$\\
&  & 7.34 (H) & & &  \\
$\Omega_{cc}^{+}\to \Xi_c^{+}\overline{K}^0$  & 1.15 & 1.36 (N) &  1.98 & \\
&  & 2.10 (H) & & &  \\
$\Omega_{cc}^{+}\to \Xi_c^{'+}\overline{K}^0$  & 0.29 & 0.61 (N) &  0.31 & \\
&  & 1.19 (H) & & &  \\
\end{tabular}
\end{ruledtabular}}
\end{center}
\end{table}

From Table \ref{tab:BF} we see that the nonfactorizable amplitudes in $\Xi_{cc}^{++}\to \Xi_c^{'+}\pi^+$  and $\Omega_{cc}^+\to \Xi_c^{'+}\overline{K}^0$ are very small compared to the factorizable ones. As stated before, the topological amplitude $C'$ in these decays should vanish due to the Pati-Woo theorem which requires that the quark pair in a baryon produced by weak interactions be antisymmetric in flavor. Since the sextet
$\Xi'_c$ is symmetric in the light quark flavor in the SU(3) limit, it cannot contribute to $C'$. It is clear from Eqs. (\ref{eq:Swave}), (\ref{eq:Xicc++Pwave}) and (\ref{eq:OmegaccPwave}) that the $C'$ amplitude is proportional to the matrix element $a_{\Xi_c^{'+}\Xi_{cc}^+}$ governed by the bag integral $X_1$ introduced in Eq. (\ref{eq:X}), which vanishes in the SU(3) limit. Likewise, the nonfactorizable $S$-wave amplitudes in $\Xi_{cc}^+\to\Xi_c^{'+}(\pi^0,\eta)$ governed by $C'$ also vanish in the limit of SU(3) symmetry. However, this is not the case for nonfactorizable $P$-wave amplitudes due to the presence of $W$-exchange contributions $E_1$ and/or $E_2$.

Finally, we notice that the two decay modes $\Xi_{cc}^{++}\to \Sigma_c^{++}\overline{K}^0$ and $\Omega_{cc}^+\to\Omega_c^0\pi^+$ are purely factorizable processes. Therefore, their theoretical calculations are much more clean. Measurements of them will provide information on $\Xi_{cc}^{++}\to \Sigma_c^{++}$ and $\Omega_{cc}^+\to\Omega_c^0$ transition form factors.  Our result of $\B(\Omega_{cc}^+\to\Omega_c^0\pi^+)\approx 4\%$ suggests that this mode may serve as a discovery channel for the $\Omega_{cc}^+$. More explicitly, it can be searched in the final state $pK^-\pi^-\pi^+\pi^+$ through the decay $\Omega_{cc}^+\to \Omega_c^0\pi^+$ followed by $\Omega_c^0\to\Omega^-\pi^+\to p\pi^- K^-\pi^+$.

\subsection{Comparison with other works}

In Table \ref{tab:Xicpi} we have already compared our calculated  partial-wave amplitudes for some of doubly charmed baryon decays with Gutsche {\it et al.} \cite{Gutsche:2018msz},  Dhir and Sharma \cite{Sharma:2017txj,Dhir:2018twm}. We agree with Gutsche {\it et al.} on the interference patterns in $S$- and $P$-wave amplitudes of $\Xi_{cc}^{++}\to \Xi_c^+\pi^+$ and $\Omega_{cc}^+\to \Xi_c^+\overline{K}^0$, but disagree on the interference patterns in $\Xi_{cc}^{++}\to \Xi_c^{'+}\pi^+$ and $\Omega_{cc}^+\to \Xi_c^{'+}\overline{K}^0$. Nevertheless, the disagreement in the last two modes is minor because of the Pati-Woo theorem  for the $C'$ amplitude. We agree with Dhir and Sharma on the interference patterns in  $P$-wave amplitudes of $\Xi_{cc}^+\to \Xi_c^{'0}\pi^+,\Xi_c^+\overline{K}^0,\Lambda_c^+\overline{K}^0$, but disagree on that in $\Xi_{cc}^{++}\to \Xi_c^+\pi^+$, $\Xi_{cc}^+\to \Xi_c^0\pi^+$ and $\Omega_{cc}^+\to \Xi_c^{+}\overline{K}^0$. Consequently, the hierarchy pattern of ${\cal B}(\Xi_{cc}^{+}\to \Xi_c^0\pi^+)$ and ${\cal B}(\Xi_{cc}^{++}\to \Xi_c^+\pi^+)$ in this work and \cite{Sharma:2017txj} is opposite to each other.

In Table \ref{tab:comparison2} we present a complete comparison of the calculated branching fractions of Cabibbo-favored $\B_{cc}\to \B_c+P$ decays with other works.
Only the factorizable contributions from the external $W$-emission governed by the Wilson coefficient $a_1$ were considered in
references \cite{Wang:2017mqp,Shi:2019hbf,Gerasimov:2019jwp,Ke:2019lcf} with nonfactorizbale effects being neglected. We see from Table \ref{tab:modes} that only the decay modes $\Xi_{cc}^{++}\to \Xi_c^{(')+}\pi^+$, $\Xi_{cc}^{+}\to \Xi_c^{(')0}\pi^+$ and $\Omega_{cc}^+\to\Omega_c^0\pi^+$ receive contributions from the external $W$-emission amplitude $T$. Branching fractions calculated in Refs. \cite{Wang:2017mqp,Ke:2019lcf,Shi:2019hbf} were based on the form-factor models LFQM(I), LFQM(II) and QSR, respectively. Since $\B_{cc}\to \B_c$ transition form factors are largest in LFQM(I) and smallest in QSR (see Table \ref{tab:FFcomp}), this leads to $\B(\Xi_{cc}^{++}\to \Xi_c^{+}\pi^+)$ and $\B(\Xi_{cc}^{+}\to \Xi_c^{0}\pi^+)$ in \cite{Shi:2019hbf} two times smaller than that in \cite{Wang:2017mqp}, for example. The authors of \cite{Gerasimov:2019jwp} employed LFQM(I) form factors, but their predictions are slightly larger than that of \cite{Wang:2017mqp}.

We see from Table \ref{tab:comparison2} that the predicted $\B(\Xi_{cc}^+\to\Xi_c^+\pi^0)$ and
$\B(\Xi_{cc}^+\to\Xi_c^+\eta)$ in \cite{Sharma:2017txj} are much smaller than ours. This is because we have sizable $W$-exchange contributions to the $S$-wave amplitudes of $\Xi_{cc}^+\to\Xi_c^+(\pi^0,\eta)$, which are absent in \cite{Sharma:2017txj}. This can be tested in the future.

\section{Conclusions}\label{sec:con}

In this work we have studied the Cabibbo-allowed decays $\B_{cc}\to \B_c+P$ of doubly charmed baryons $\Xi_{cc}^{++}, \Xi_{cc}^+$ and $\Omega_{cc}^+$.
To estimate the nonfactorizable contributions, we work in the pole model for the $P$-wave amplitudes and current algebra for $S$-wave ones.
Throughout the whole calculations, all the non-perturbative parameters including form factors, baryon matrix elements and
axial-vector form factors are evaluated within the framework of the MIT bag model.

We draw some conclusions from our analysis:
\begin{itemize}

\item All the unknown parameters such as $\B_{cc}\to\B_c$ transition form factors, the matrix elements $a_{\B'\B}$ and the axial-vector form factors $g^{A(P)}_{\mathcal{B}'\mathcal{B}}$ are evaluated in the same MIT bag model to ensure the correctness of their relative signs once the wave function convention is fixed.
    For the $\Xi_{cc}^{++}\to \Xi_c^+\pi^+$ mode, we found a large destructive interference between factorizable and nonfactorizable contributions for both $S$- and $P$-wave amplitudes. Our prediction of $\sim 0.70\%$ for its branching fraction is smaller than the earlier estimates in which nonfactorizable effects were not considered but agrees nicely with the result based on an entirely different approach, namely, the covariant confined quark model. On the contrary, a large constructive interference was found in the $P$-wave amplitude by Sharma and Dhir \cite{Sharma:2017txj}, leading to a branching fraction of order $(7-16)\%$. For example,
    it is the relative sign between the form factor $g_1$ and the combination $a_{\Xi_c^+\Xi_{cc}^+}\times g_{\Xi_{cc}^{++}\Xi_{cc}^+\pi^+}$ that accounts for the different $P$-wave interference pattern in $\Xi_{cc}^{++}\to \Xi_c^+\pi^+$ found in this work and the work by Sharma and Dhir.

\item Using the current results of the absolute branching fractions of $(\Lambda_c^+,\Xi_c^+)\to p K^-\pi^+$ and the LHCb measurement of $\Xi_{cc}^{++}\to\Xi_c^+\pi^+$ relative to $\Xi_{cc}^{++}\to\Lambda_c^+ K^- \pi^+\pi^+$,
we obtain $\B(\Xi_{cc}^{++}\to\Xi_c^+\pi^+)_{\rm expt}\approx (1.83\pm1.01)\%$ after employing the latest prediction of $\B(\Xi_{cc}^{++}\to\Sigma_c^{++}\overline{K}^{*0})$ and the plausible assumption of $\B(\Xi_{cc}^{++}\to\Lambda_c^+ K^- \pi^+\pi^+)\approx {2\over 3}\B(\Xi_{cc}^{++}\to\Sigma_c^{++}\overline{K}^{*0})$. Therefore, our prediction of $\mathcal{B}(\Xi_{cc}^{++}\to\Xi_c^+\pi^+)\approx 0.7\%$ is consistent with the experimental value but in the lower end. It is important to pin down the branching fraction of $\Xi_{cc}^{++}\to\Xi_c^+\pi^+$ in future study.

\item
Factorizable and nonfactorizable $S$-wave amplitudes interfere constructively in $\Xi_{cc}^+\to\Xi_c^0\pi^+$. Its large branching fraction of order 4\% may enable experimentalists to search for the $\Xi_{cc}^+$ through this mode. In this way, $\Xi_{cc}^+$ is reconstructed through the $\Xi_{cc}^+\to\Xi_c^0\pi^+$ followed by the decay chain $\Xi_c^0\to \Xi^-\pi^+\to p\pi^-\pi^-\pi^+$.

\item Besides $\Xi_{cc}^+\to\Xi_c^0\pi^+$, the $\Xi_{cc}^+\to\Xi_c^+ (\pi^0,\eta)$ modes also receive large nonfactorizable contributions to their $S$-wave amplitudes. Hence, they have large branching fractions among $\Xi_{cc}^+\to \B_c+P$ decays.

\item
The two decay modes $\Xi_{cc}^{++}\to \Sigma_c^{++}\overline{K}^0$ and $\Omega_{cc}^+\to\Omega_c^0\pi^+$ are purely factorizable processes.  Measurements of them will provide information on $\Xi_{cc}^{++}\to \Sigma_c^{++}$ and $\Omega_{cc}^+\to\Omega_c^0$ transition form factors.  Our calculation of $\B(\Omega_{cc}^+\to\Omega_c^0\pi^+)\approx 4\%$ suggests that this mode may serve as a discovery channel for the $\Omega_{cc}^+$. That is, it can be searched in the final state $pK^-\pi^-\pi^+\pi^+$ through the decay $\Omega_{cc}^+\to \Omega_c^0\pi^+$ followed by $\Omega_c^0\to\Omega^-\pi^+\to p\pi^- K^-\pi^+$.

\item
Nonfactorizable amplitudes in $\Xi_{cc}^{++}\to \Xi_c^{'+}\pi^+$  and $\Omega_{cc}^+\to \Xi_c^{'+}\overline{K}^0$ are very small compared to the factorizable ones owing to the Pati-Woo theorem for the inner $W$-emission amplitude. Likewise, nonfactorizable $S$-wave amplitudes in $\Xi_{cc}^+\to\Xi_c^{'+}(\pi^0,\eta)$ decays are also suppressed by the same mechanism.

\end{itemize}

\vskip 2 cm
\begin{acknowledgments}
We are grateful to Rohit Dhir for discussions.
This research was supported in part by the Ministry of Science and Technology of R.O.C. under Grant No. 107-2119-M-001-034. F. Xu is supported by NSFC under Grant Nos. U1932104 and 11605076.
\end{acknowledgments}

\appendix

\section{Wave functions of doubly charmed baryons}

Throughout the whole calculation, baryon wave functions
are adopted from the convention in \cite{Cheng:2018hwl}.
Here we add the wave functions of doubly charmed baryons with $S_z=1/2$:
\begin{align}
& \Xi_{cc}^{++}=-\frac{1}{\sqrt{3}}\left[ ccu\chi_s +(23)+(13)\right],\qquad
\Xi_{cc}^{+}=-\frac{1}{\sqrt{3}}\left[ ccd\chi_s +(23)+(13)\right],  \nonumber \\
& \hspace{3cm} \Omega_{cc}^{+}=-\frac{1}{\sqrt{3}}\left[ ccs\chi_s +(23)+(13)\right],
\end{align}
where $abc\chi_{_S}=(2a^\up b^\up c^\dw-a^\up b^\dw c^\up-a^\dw b^\up c^\up)/\sqrt{6}$.


\end{document}